\documentclass[xcolor=tex, dvipsnames]{article}

\usepackage[utf8]{inputenc}
\usepackage{microtype}
\usepackage{graphicx}

\usepackage{hyperref}

\usepackage[accepted]{mlsys2022}

\usepackage{booktabs}   
\usepackage{subcaption} 

\usepackage{tikz}
\usepackage{tikz}
\usetikzlibrary{positioning, arrows.meta}
\usetikzlibrary{decorations.pathreplacing}
\tikzstyle{mygrid}=[cmcolor, draw=gray!40]
\usepackage{colortbl}
\usepackage{amsmath}
\usepackage{amsthm}
\usepackage{amssymb}
\makeatletter
\newcommand{\ostar}{\mathbin{\mathpalette\make@circled\star}}
\newcommand{\make@circled}[2]{%
  \ooalign{$\m@th#1\smallbigcirc{#1}$\cr\hidewidth$\m@th#1#2$\hidewidth\cr}%
}
\newcommand{\smallbigcirc}[1]{%
  \vcenter{\hbox{\scalebox{0.77778}{$\m@th#1\bigcirc$}}}%
}
\makeatother

\usepackage{ottalt}

\newcommand{\ottafterlastrule}{\\}

  \renewottcommands[ott]

\renewcommand\ottaltinferrule[4]{
  \inferrule*[narrower=1,lab=#1,#2]
    {#3}
    {#4}
}

\usepackage{cleveref}
\usepackage{float}
\usepackage{xspace}
\usepackage{color, colortbl}
\usepackage{hhline}

\usepackage{subcaption}
\usepackage{amsfonts} 
\usepackage{multirow}

\definecolor{cmcolor}{RGB}{41,128,185}
\definecolor{rowcolor}{RGB}{227,244,249} 
\definecolor{headrowcolor}{RGB}{73,203,219}
\newcommand{\makecmcolor}[1]{{\textcolor{cmcolor}{#1}}}
\newcommand*\circled[1]{\tikz[baseline=(char.base)]{
            \node[shape=circle,draw,inner sep=0.5pt] (char) {\small{{#1}}};}}

\usepackage{caption}
\usepackage{relsize}
\usepackage{comment}
\newcolumntype{P}[1]{>{\centering\arraybackslash}p{#1}}
\newcolumntype{M}[1]{>{\centering\arraybackslash}m{#1}}

\usepackage{longtable}

\newtheorem{theorem}{Theorem}[section]
\newtheorem{definition}[theorem]{Definition}
\newtheorem{lemma}[theorem]{Lemma}
\newtheorem{corollary}[theorem]{Corollary}

\newcommand{\reducescatter}{\mathsf{ReduceScatter}}
\newcommand{\allreduce}{\mathsf{AllReduce}}
\newcommand{\allgather}{\mathsf{AllGather}}

\newcommand{\reduceno}{\mathsf{Reduce}}
\newcommand{\broadcastno}{\mathsf{Broadcast}}

\newcommand{\hoarenoline}[3]{\{\,#1\,\}\,#2\,\{\,#3\,\}}
\newcommand{\disjoint}{\ostar}
\newcommand{\add}{\uplus}
\newcommand{\union}{\add}

\newcommand{\G}{\mathcal{G}}
\newcommand{\op}{\mathcal{C}}

\newcommand{\insidegroup}{\mathsf{InsideGroup}}
\renewcommand{\parallel}[1]{\mathsf{Parallel}({#1})}
\newcommand{\master}[1]{\mathsf{Master}({#1})}
\newcommand{\parallelno}{\mathsf{Parallel}}
\newcommand{\masterno}{\mathsf{Master}}

\newcommand{\slice}{slice}
\newcommand{\form}{form}
\newcommand{\reduction}{reduction}
\newcommand{\program}{program}

\newcommand{\group}[1]{\{#1\}}

\newcommand{\bool}{\mathbb{B}}

\newcommand{\mynote}[3]{{\itshape\textcolor{#2}{\textbf{#1:} #3}}}
\renewcommand{\mynote}[3]{{}}  
\newcommand{\ningning}[1]{\mynote{Ningning}{purple}{#1}}
\newcommand{\dimitrios}[1]{\mynote{Dimitrios}{blue}{#1}}

\newcommand{\toolname}{$P^{2}$\xspace}
\newcommand{\axes}[1]{{\small{$\begin{bmatrix} #1 \end{bmatrix}$}}}
\newcommand{\axesinside}[1]{{\small{\begin{bmatrix} #1 \end{bmatrix}}}}

\usepackage{microtype}

\usepackage{thm-restate}
\usepackage{bm}

\begin{document}

\twocolumn[
\mlsystitle{Synthesizing Optimal Parallelism Placement and Reduction Strategies on Hierarchical Systems for Deep Learning}



\mlsyssetsymbol{equal}{*}

\begin{mlsysauthorlist}
\mlsysauthor{Ningning Xie}{cam}
\mlsysauthor{Tamara Norman}{dm}
\mlsysauthor{Dominik Grewe}{dm}
\mlsysauthor{Dimitrios Vytiniotis}{dm} \end{mlsysauthorlist}

\mlsysaffiliation{cam}{University of Cambridge}
\mlsysaffiliation{dm}{DeepMind}

\mlsyscorrespondingauthor{Ningning Xie}{nx213@cam.ac.uk}

\mlsyskeywords{Machine Learning, MLSys}

\vskip 0.3in

\begin{abstract}
We present a novel characterization of the mapping of multiple parallelism forms (e.g. data and model parallelism) onto hierarchical accelerator systems that is hierarchy-aware and
greatly reduces the space of software-to-hardware mapping. We experimentally verify the substantial effect of
these mappings on all-reduce performance (up to 448$\times$). We offer a novel syntax-guided program synthesis framework that is able to decompose reductions over one or more parallelism axes to sequences of collectives in a hierarchy- and mapping-aware way. For 69\% of parallelism placements and user requested reductions, our framework synthesizes programs that outperform the default all-reduce implementation when evaluated on different GPU hierarchies (max 2.04$\times$, average 1.27$\times$). We complement our synthesis tool with a simulator exceeding 90\% top-10 accuracy, which therefore reduces the need for massive evaluations of synthesis results to determine a small set of optimal programs and mappings.
\end{abstract}
] 



\printAffiliationsAndNotice{}  

\section{Introduction}  

To facilitate efficient training of large-scale deep learning models,
numerous parallelism techniques
have been successfully employed.
Common forms of parallelism include
\textit{data parallelism}~\cite{krizhevsky2012imagenet}, where each device has a copy of the full
model to process a portion of the training data,
and \textit{model parallelism}~\cite{dean2012large}, which partitions a training model over available devices, such as
\textit{parameter sharding}~\cite{shoeybi2020megatronlm} and \textit{pipeline parallelism}~\cite{huang2019gpipe}.
More recent studies explore combinations of
parallelism forms to maximize training throughput~\cite{jia2019beyond, narayanan2021efficient},
where each form of parallelism is referred to as a \textit{parallelism axis}.

While the aforementioned forms of parallelism and their combinations have greatly improved training throughput,
they may still incur significant \textit{communication cost}.
For example, in the simplest form of data parallelism, parameter gradients for each device must be reduced and replicated for each iteration~\cite{amodei2016deep},
which is typically implemented using the \textit{collective operation} $\allreduce$~\cite{thakur2005optimization}.
State-of-the-art parameter sharding for transformers~\cite{shoeybi2020megatronlm} introduces sharded layers where each involves several $\allreduce$ operations.
Communication overhead is especially important for distributed deep learning,
as the more devices we have, computation time reduces, and the communication cost becomes more
prominent~\cite{sergeev2018horovod,goyal2018accurate}.

\begin{figure}
\small
    \centering
    \begin{subfigure}[b]{0.36\linewidth}
    \centering
    \includegraphics[width=1\textwidth]{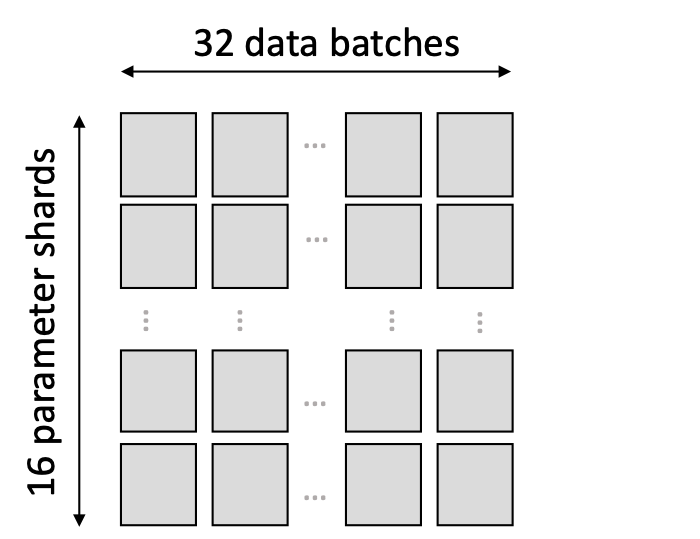}
    \caption{Combining\\parameter sharding\\ and data parallelism}
    \label{fig:intro:1}
    \end{subfigure}
    \begin{subfigure}[b]{0.3\linewidth}
    \centering
    \includegraphics[width=0.7\textwidth]{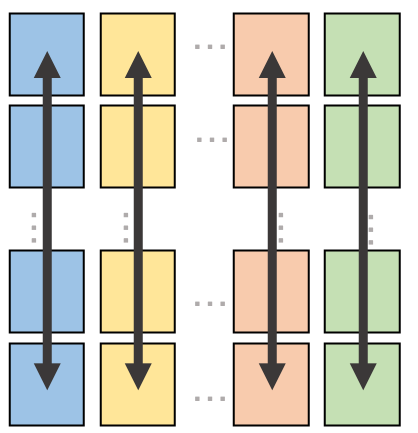}
    \caption{Reduction\\ along the axis of\\ parameter sharding}
    \label{fig:intro:2}
    \end{subfigure}
    \begin{subfigure}[b]{0.3\linewidth}
    \centering
    \includegraphics[width=0.7\textwidth]{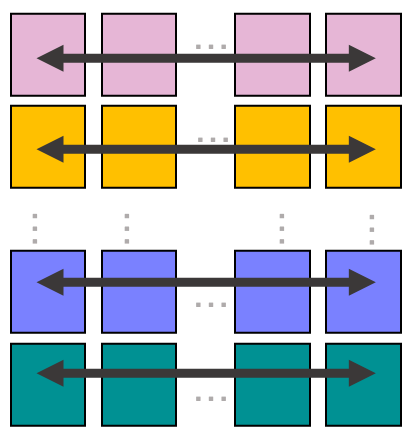}
    \caption{Reduction\\ along the axis of\\ data parallelism}
    \label{fig:intro:3}
    \end{subfigure}
    \caption{Parallelism combination}
    \label{fig:my_label}
\end{figure}

To reduce communication overhead,
one particular challenge posed by multiple parallelism axes is \textit{parallelism placement}.
That is, how we map parallelism over devices decides which devices communicate with each other along each parallelism axis, and therefore decides the communication overhead.
For example, \Cref{fig:intro:1} presents a combination of parameter sharding and data parallelism,
for which reduction along the axis of parameter sharding (or data parallelism), referred to as the \textit{reduction axis},
is shown in \Cref{fig:intro:2} (or \Cref{fig:intro:3}, respectively).
Now, suppose we map each box in the figure to devices.
In that case, different mappings correspond to different reduction device groups, which can have
a significant impact on the communication overhead depending on the network topology.

\begin{figure*}
    \centering
    \begin{subfigure}[b]{0.33\textwidth}
    \centering
    \includegraphics[width=0.7\linewidth]{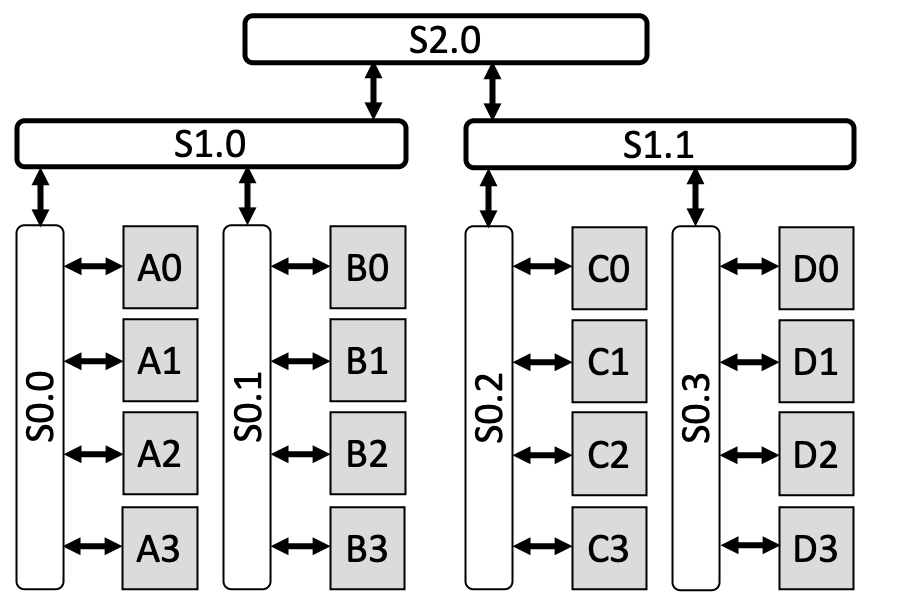}
    \caption{
    [(rack, 1), (server, 2), (CPU, 2), (GPU, 4)]
    }
    \label{fig:overview:1}
    \end{subfigure}
    \begin{subfigure}[b]{0.2\textwidth}
    \centering
    \includegraphics[width=0.6\linewidth]{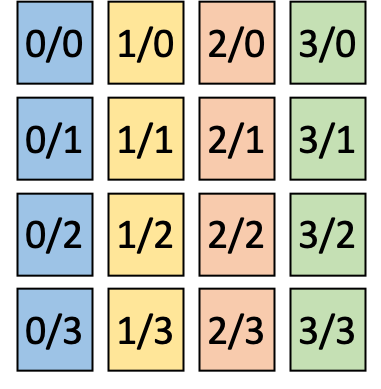}
    \caption{\scriptsize{$\begin{bmatrix}1 & 2 & 2 & 1\\ 1 & 1 & 1& 4\end{bmatrix}$}}
    \label{fig:overview:2}
    \end{subfigure}
    \begin{subfigure}[b]{0.2\textwidth}
    \centering
    \includegraphics[width=0.6\linewidth]{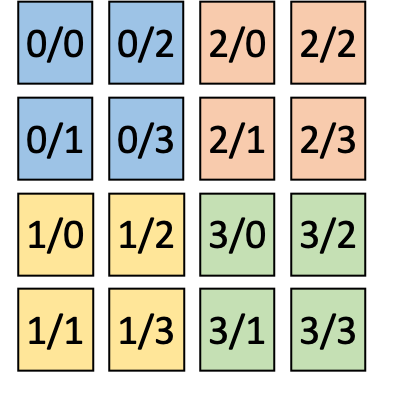}
    \caption{\scriptsize{$\begin{bmatrix}1 & 2 & 1 & 2\\ 1 & 1& 2 & 2\end{bmatrix}$}}
    \label{fig:overview:3}
    \end{subfigure}
    \begin{subfigure}[b]{0.2\textwidth}
    \centering
    \includegraphics[width=0.6\linewidth]{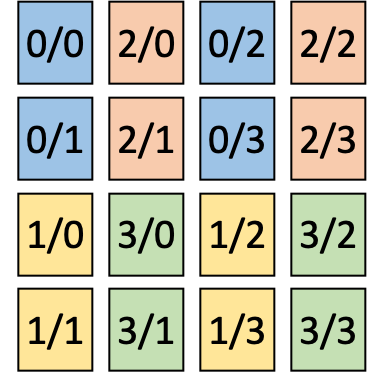}
    \caption{\scriptsize{$\begin{bmatrix}1 & 1 & 2 & 2\\ 1 & 2 & 1 & 2\end{bmatrix}$}}
    \label{fig:overview:4}
    \end{subfigure}
    \caption{(a): A system. (b), (c), (d): Possible (non-exhaustive) parallelism placements
    for (a) under data parallelism of size 4 and 4 parameter shards.
    For clarity, we show only the 16 GPUs but omit interconnects. 
    Device marker n/m indicates data batch n and parameter shard m.}
    \label{fig:overview}
\end{figure*}

In this work, we present \toolname, a tool for parallelism placement and placement-aware
synthesis of recduction strategies. In particular, we offer the following contributions:

\begin{itemize}
    \item \textit{Parallelism placement synthesis}: Given the parallelism axes, the reduction axes,
    and a hierarchical system
    topology, \toolname automatically
    synthesizes \textit{hierarchical} parallelism placements,
    where a parallelism placement is modelled as a \textit{parallelism matrix} mapping
    from parallelism axes to the system hierarchy (\Cref{sec:parallel-placement}).
    The notion of parallelism matrices
    greatly reduces the space of parallelism placements
    contrary to a naive implementation.
    
    \item \textit{Reduction strategy synthesis}: For each parallelism placement, \toolname utilizes
    the system hierarchy to further synthesize a wide variety of \textit{reduction strategies} to implement
    reductions using common collective operations. To achieve this, 
    we introduce: (a)
    a formal semantics
    for collectives (\Cref{sec:collective-operations}) 
    based on \textit{Hoare} triples~\cite{hoare};
    (2) a domain-specific language (DSL) that can express
    possibly simultaneous reductions amongst groups of devices based on
    the system hierarchy (\Cref{sec:reduction-programs}); and
    (b) a lowering of our DSL into sequences of
    collective operations.
    We use the formal semantics to guide a syntax-directed synthesis procedure 
    on our DSL.
    
    \item \textit{Synthesis hierarchy}: 
    We show how the parallelism matrix,
    which determines a candidate parallelism placement,
    can be put to good use by the synthesizer to massively reduce the space of programs considered
    {\em without missing any semantically valid} programs -- provably (\Cref{sec:system-hierarchy}).

    \item \textit{Evaluation}: We evaluate the parallelism matrices and reduction strategies synthesized
    by \toolname on two different GPU systems available on Google Cloud Platform (GCP) (\Cref{sec:evaluation}).
    We use 
    collective operations as implemented by NVIDIA's NCCL communication library~\cite{nccl}, exposed through XLA.
    The evaluation demonstrates (1) the impact
    of parallelism placement: the performance
    of a single $\allreduce$ across different parallelism matrices differs up to 448.5$\times$;
    and (2) the effectiveness of custom reduction strategies: 
    for 69\% of all parallelism mapping matrices,
    a synthesized reduction outperforms $\allreduce$ with up to 2.04$\times$
    speedup (average 1.27$\times$).
    
    \item \textit{Simulation}: \toolname synthesizes all 
    mapping and hierarchy-aware reduction strategies, but evaluating
    hundreds or thousands of them to identify the best can be expensive.
    We therefore introduce
    a simulator for predicting the end-to-end performance of a parallelism
    matrix and reduction strategy (\Cref{sec:simulation}). The simulator 
    is aware of the network topology including different bandwidths for different interconnects and networks
    (e.g., NVLink and ethernet / data-center network in GPU topologies),
    predicting with reasonable accuracy the 
    communication overhead for each parallelism placement
    and reduction strategy.
    The validation -- over all mappings and synthesized programs for each mapping, and for each of the two GPU systems we considered --  demonstrates that
    the simulator has 52\%, 72\%, and 92\% of top-1, top-5 and top-10 accuracy, respectively, making it practical for identifying a much smaller subset
    of programs for actual evaluation.
\end{itemize}

\toolname is helpful for ML practitioners to speed up their models by
improving placement and synthesizing reduction strategies tailored
to their system hierarchies. For instance, we have used \toolname to 
improve Resnet-50~\cite{resnet50} data-parallel training by 15\% across 
4 nodes, each with 8 V100 GPUs. (See~\Cref{sec:evaluation} for the details of this
system.)

\section{Overview}
\label{sec:overview}

This section outlines the key design in \toolname.
First, a \textit{system} consists of two entities: (1) a hardware
hierarchy, where each
level has a name and a cardinality; and (2) 
a set of switched interconnects.
The system hierarchy is expected to reflect how devices are arranged.
\Cref{fig:overview:1} describes an example system with 16 GPUs~\cite{cho2019blueconnect}.
The hierarchy is one-dimensional: a rack has 2 servers,
each with 2 CPUs 
connecting 4 GPUs.
Interconnects specify
how devices are connected with each other 
and the latency and bandwidth constraints.
In this case, we have exactly one kind of interconnect in each
level,
but, in general, the interconnect topology can be more complex:
there can be multiple interconnects in one level, and
an interconnect can connect devices (and other interconnects) across levels.

\subsection{Parallelism Placement}
\label{sec:overview:pp}

Parallel placement decides which parts of a partitioned program
will execute on which parts of a system.
However, synthesizing all arbitrary device mappings,
as well as running experiments with them,
can be extremely expensive if implemented naively.
For example,
if we have data parallelism of size 4 
and 4 parameter shards for the system in \Cref{fig:overview:1},
then there will be $(4*4)! > 2^{44}$ possibilities
to decide which partitioned program maps to which GPU.

To explore the search space efficiently,
the critical idea of 
\toolname is to \textit{partition parallelism axes over the system hierarchy}
to generate topology-aware parallelism placements,
while still being able to systematically generate a wide range of parallelism placements.
Specifically, a result of parallelism placement synthesis is a \textit{parallelism matrix},
where each element is a
\textit{parallelism factor} representing the number of a specific level
in the hierarchy that 
a parallelism form splits the computation across.
Figures~\ref{fig:overview:2}, \ref{fig:overview:3} and \ref{fig:overview:4}
show examples of parallelism matrices synthesized by \toolname,
where
we have data parallelism of size 4 and 4 parameter shards.
In \Cref{fig:overview:2},
the first row {\small{$\begin{bmatrix}1 & 2 & 2 & 1\end{bmatrix}$}}
corresponds to a factorization of data parallelism on each system level.
Specifically,
we first assign all data parallelism (of size $4$) into 1 rack (each with
data parallelism of size $4/1 = 4$).
Then each rack assigns
data parallelism of size $4$ into 2 servers
(each with data parallelism of size $4/2 = 2$).
Next, each server
assigns data parallelism of size $2$
into 2 CPUs (each with data parallelism of size $2/2=1$).
Finally, each CPU assigns data parallelism of size $1$ into 1 GPU.
The second row {\small{$\begin{bmatrix}1 & 1 & 1 & 4\end{bmatrix}$}}
corresponds to a factorization of parameter sharding:
each rack, server, and CPU gets assigned all parameter shards (of size 4),
and each CPU then assigns $4$ parameter shards into 4 GPUs,
each GPU level with $4/4 = 1$ shard.
Therefore, 
in the resulting placement,
each CPU corresponds to one replica (data parallelism)
where each GPU has one parameter shard.
We can interpret
\Cref{fig:overview:3} and \ref{fig:overview:4} accordingly.

Note how parallelism matrices decide communication requirements.
Consider reduction along parameter sharding (i.e., reduce devices n/m with the same n but different m).
In \Cref{fig:overview:2}, this can be done by communication over only S0,
while in \ref{fig:overview:3}, half of the data can be reduced by only S0, but the
rest of the reduction requires
communication over S0/S1/S2.
We discuss the impact of parallelism placements on communication cost in detail in
\Cref{sec:evaluation}.

\begin{figure}
    %
     \begin{subfigure}[t]{0.09\textwidth}
    \centering
    \includegraphics[width=0.7\linewidth]{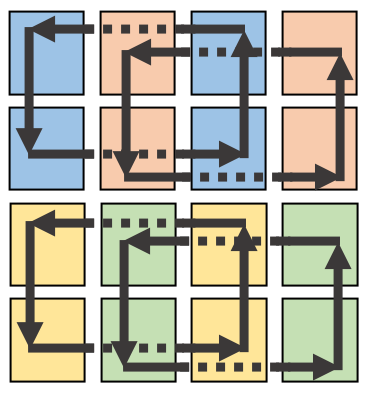}
    \caption{\scriptsize{$\allreduce$}}
    \label{fig:reduction:0}
    \end{subfigure}
    \begin{subfigure}[t]{0.17\textwidth}
    \centering
    \includegraphics[width=0.36\linewidth]{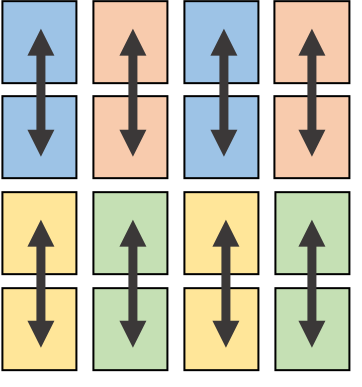}
    \includegraphics[width=0.37\linewidth]{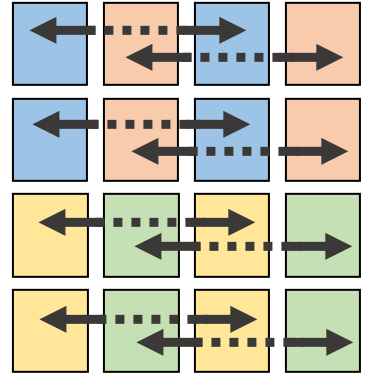}
    \caption{\scriptsize{$\allreduce$-$\allreduce$}}
    \label{fig:reduction:1}
    \end{subfigure}
    \begin{subfigure}[t]{0.21\textwidth}
    \centering
    \includegraphics[width=0.27\linewidth]{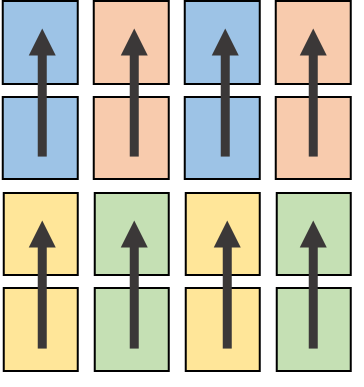}
    \includegraphics[width=0.28\linewidth]{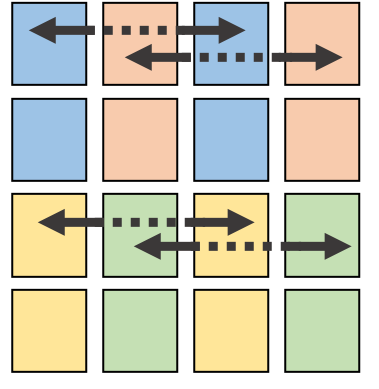}
    \includegraphics[width=0.27\linewidth]{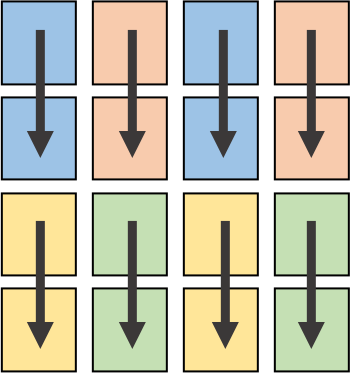}
    \caption{\scriptsize{$\reduceno$-$\allreduce$-$\broadcastno$}}
    \label{fig:reduction:2}
    \end{subfigure}
    \caption{Example reduction strategies.}
    \label{fig:reduction}
\end{figure}

\begin{figure}
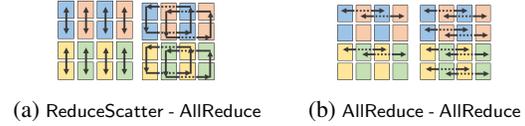

    \centering
    \begin{subfigure}[t]{0.21\textwidth}
    \centering
    \includegraphics[width=0.28\linewidth]{images/reduction5.png}
    \includegraphics[width=0.28\linewidth]{images/reduction1.png}
    \caption{\scriptsize{$\reducescatter$ - $\allreduce$}}
    \label{fig:reduction:false:1}
    \end{subfigure}
    \begin{subfigure}[t]{0.21\textwidth}
    \centering
    \includegraphics[width=0.28\linewidth]{images/reduction3.png}
    \includegraphics[width=0.28\linewidth]{images/reduction6.png}
    \caption{\scriptsize{$\allreduce$ - $\allreduce$}}
    \label{fig:reduction:false:2}
    \end{subfigure}
    \caption{Semantically invalid reduction.
    (a): Reduce data that should not be reduced.
    (b): Reduce the same data twice.}
    \label{fig:reduction:false}
\end{figure}
\subsection{Reduction Strategy}
For each parallelism matrix, \toolname further synthesizes
topology-aware reduction strategies
using common collective operations,
which allows us to find the optimal reduction strategy
for any given parallelism matrix.

To illustrate the idea, consider the parallelism matrix in
\Cref{fig:overview:4}, and the goal is to reduce along
parameter sharding.
As shown in \Cref{fig:reduction:0},
an obvious choice to perform the reduction is a single $\allreduce$ within reduction groups.
However, such reduction may be suboptimal,
as it does not utilize the topology of the system.
\Cref{fig:reduction:1} and \ref{fig:reduction:2} show two reduction strategies, among others, synthesized by \toolname.
\Cref{fig:reduction:1} first performs a step of $\allreduce$ which communicates over only S0,
and then $\allreduce$ that communicates over S0/S1/S2.
\Cref{fig:reduction:2} first performs $\reduceno$ that puts the reduction result in the root device,
then $\allreduce$ between root devices,
and finally $\broadcastno$ that broadcasts data from the root device.
Of particular interest in these two reduction strategies is that no
one is strictly better than the other, as the communication overhead depends
on the network: \ref{fig:reduction:2} takes more steps, but has fewer data
to be transferred over S1/S2,
which may outperform \ref{fig:reduction:1} if S0 has high bandwidth while
communication over S1/S2 is expensive.

\toolname gives us a systematic way to synthesize and compare a wide range of
topology-aware reduction strategies.
In particular, synthesized reduction strategies can outperform a single step
$\allreduce$, with speedup up to 2.05$\times$.
However, synthesizing reduction strategies also imposes challenges, which we outline in the rest of this section.

\subsection{Formalism of Collective Operations}
\label{sec:overview:collective:operations}

To synthesize reduction strategies, we first need to formalize the semantics of 
collective operations, since not all sequences of 
operationally
valid collective operations correspond to semantically correct implementations of
the end-to-end reduction requested by the user.
For example, consider the (incomplete) reduction steps given in \Cref{fig:reduction:false}
for the requested reduction across
parameter shards.
Both programs can be executed successfully, e.g., by NCCL~\cite{nccl}.
Unfortunately, they are both \textit{semantically invalid}.
In particular,
\textit{we consider reduction steps which result in
device states that can never reach the final desired state to be semantically invalid}.
Specifically,
in \ref{fig:reduction:false:1},
the first $\reducescatter$ will reduce, among others,
device A0 and A1 (recall that GPUs are named in \Cref{fig:overview:1}),
and put the first half of the result on A0 and the second half on A1.
Then the second $\allreduce$ will reduce A0 and A1 -- so the first and the second half of
the result get reduced while they should not! 
Now, we can never reach the desired final state.
\ref{fig:reduction:false:2} is also
invalid as it reduces the data on A0 and C0 twice.

\toolname provides a concise and novel formalism of 
common collective operations (\Cref{sec:collective-operations}) 
that captures semantic correctness
and rules out semantically invalid programs, massively reducing the synthesis space.
Specifically,
each device state is defined as a \textit{state matrix} describing
what kind of data a device has.
The semantics of collective operations is defined with Hoare triples~\cite{hoare},
where a collective takes the state of each device as a \textit{pre-condition} and returns a
new state as a \textit{post-condition}.

\subsection{Reduction Communication Patterns}

Even though the formalism of collective operations rules out semantically invalid reduction steps,
the search space of reduction strategies
is still quite large. One reason is that
we need to decide which devices form a reduction group
for each reduction step.
For example, the first step in \Cref{fig:reduction:1}
reduces over $\group{A0, A1}, \group{A2, A3}$ (among others).
We may randomly generate all possible groups,
but that would significantly increase the search space. Also,
many of them
would be immediately thrown away after semantic checks.

To synthesize reduction strategies effectively,
\toolname uses a domain-specific language (\Cref{sec:reduction-programs})
that explores the hierarchy to
generate \textit{hierarchical} communication patterns.
The reduction language, together with the \textit{synthesis hierarchy} (explained next),
can model many common communication patterns,
including those in \Cref{fig:reduction} and \ref{fig:reduction:false}.

\subsection{Synthesis Hierarchy}
\label{sec:overview:hierarchy}

\begin{table}
    \begin{tabular}{llll} \toprule
         \multirow{3}{*}{{\small{$\begin{bmatrix}1 & 1 & 2 & 2 \\ \makecmcolor{1} & \makecmcolor{2} & \makecmcolor{1} & \makecmcolor{2} \end{bmatrix}$}}}
           & column-based & {\small{$\begin{bmatrix} 1 & \makecmcolor{1} & 1 & \makecmcolor{2} & 2 & \makecmcolor{1} & 2 & \makecmcolor{2}\end{bmatrix}$}} &\circled{1}\\
          & row-based & {\small{$\begin{bmatrix}1 & 1 & 2 & 2 & \makecmcolor{1} & \makecmcolor{2} & \makecmcolor{1} & \makecmcolor{2} \end{bmatrix}$}} &\circled{2} \\
          & reduction axis &  {\small{$\begin{bmatrix}\makecmcolor{1} & \makecmcolor{2} & \makecmcolor{1} & \makecmcolor{2}\end{bmatrix}$}} &\circled{3}\\ \hline
         \multirow{4}{*}{{\small{$\begin{bmatrix} \makecmcolor{1} & \makecmcolor{2} & \makecmcolor{3}\\4 & 5 & 6 \\ \makecmcolor{7} & \makecmcolor{8} & \makecmcolor{9}\end{bmatrix}$}}}
           & column-based & {\small{$\begin{bmatrix} \makecmcolor{1} & 4 & \makecmcolor{7} & \makecmcolor{2} & 5 & \makecmcolor{8} & \makecmcolor{3} & 6  & \makecmcolor{9}\end{bmatrix}$}} & \\
          & row-based & {\small{$\begin{bmatrix} \makecmcolor{1} & \makecmcolor{2} & \makecmcolor{3} & 4 & 5 & 6 & \makecmcolor{7}  & \makecmcolor{8}  & \makecmcolor{9}\end{bmatrix}$}} & \\
          & reduction axes &  {\small{$\begin{bmatrix}\makecmcolor{1} & \makecmcolor{2} & \makecmcolor{3} & \makecmcolor{7} & \makecmcolor{8} & \makecmcolor{9} \end{bmatrix}$}} \\
          & collapsed & {\small{$\begin{bmatrix}\makecmcolor{7} & \makecmcolor{16} & \makecmcolor{27} \end{bmatrix}$}} \\
          \bottomrule
    \end{tabular}
    \caption{Synthesis hierarchy (reduction axes highlighted)}
    \label{fig:hierarchies}
\end{table}

\begin{figure}
\begin{minipage}[t]{0.07\textwidth}
    \includegraphics[width=0.9\textwidth]{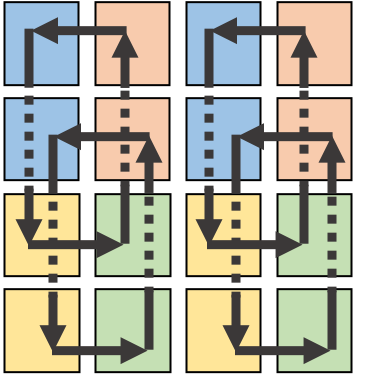}
    \caption{}
    \label{fig:global-pattern}
\end{minipage}
\begin{minipage}{0.03\textwidth}
\qquad
\end{minipage}
\begin{minipage}[t]{0.4\textwidth}
    \includegraphics[width=0.82\linewidth]{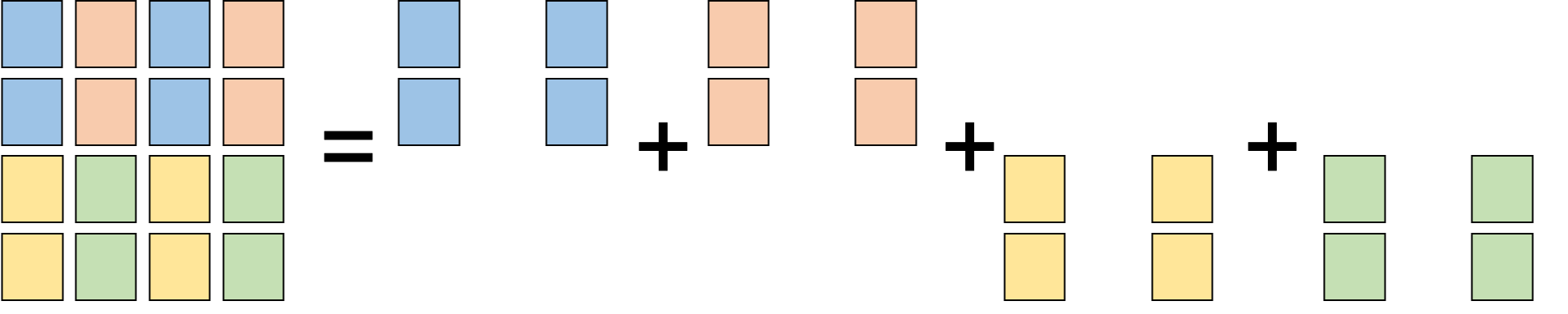}
    \caption{Device structured as the reduction axis}
    \label{fig:composition}
\end{minipage}
\end{figure}

To generate 
hierarchical reduction communication patterns,
the DSL reduction instruction needs to know the synthesis hierarchy.
For example, a possible instruction is to
reduce all "GPUs" connected to the same "CPU".
Now,
an important design decision to consider
is \textit{which} hierarchy to use for synthesis, as it decides what kind of instructions we can produce.
One obvious choice is the hardware hierarchy,
i.e., \axes{1 & 2 & 2 &4} for our running example
(we ignore level names as that can be randomly generated).
But the system hierarchy is not fine-grained enough for the requested reduction:
for example, the reduction in \Cref{fig:reduction} requires
reducing half of the GPUs connected to a CPU.
To do that, we need to take parallelism axes into consideration.
In this case, the parallelism matrix is
{\small{$\begin{bmatrix}1 & 1 & 2 & 2 \\ \makecmcolor{1} & \makecmcolor{2} & \makecmcolor{1} & \makecmcolor{2} \end{bmatrix}$}},
which splits GPU into 2 by $\makecmcolor{2}$,
allowing us to reduce $\makecmcolor{2}$ GPUs connected to a CPU.
However, to form a synthesis hierarchy from
the parallelism matrix, we have two options
(\Cref{fig:hierarchies}):
\circled{1}
puts parallelism factors by columns, which essentially
expands the system hierarchy corresponding to the parallelism matrix;
or
\circled{2}
puts factors by rows,
which expands the parallelism axes.
In this work,
we \textit{prove} that 
\circled{2} is more \textit{expressive}
than \circled{1},
i.e., \circled{2} can generate all semantically valid reduction strategies
that can be generated by \circled{1}.
The result might be somewhat counter-intuitive,
as \circled{1} seems a natural way to expand the
system hierarchy.
The critical insight is that
as \circled{2} puts parallelism axes consecutively,
it can more easily generate semantically correct reduction,
while
\circled{1}
can more easily reduce devices laid out consequently but
those reduction can be semantically invalid as it partitions
parallelism axes.

It turns out we can further optimize the synthesis hierarchy. In particular, note that
we reduce along the
axis {\small{ $\begin{bmatrix}\makecmcolor{1} & \makecmcolor{2} & \makecmcolor{1} & \makecmcolor{2}\end{bmatrix}$}},
but \circled{2} includes the full
matrix (i.e., including \axes{1 & 1 & 2 & 2}).
With a full matrix,
we can generate reduction like \Cref{fig:global-pattern},
which, however,
is not useful for this specific case, as we should not reduce device A0 and A2.
The key observation here is that each reduction group is
essentially structured according to the reduction axis
{\small{$\begin{bmatrix}\makecmcolor{1} & \makecmcolor{2} & \makecmcolor{1} & \makecmcolor{2}\end{bmatrix}$}},
and this structure is repeated for the rest of the matrix, as shown in \Cref{fig:composition}.
Based on this observation, \toolname uses the synthesis hierarchy \circled{3} formed by
parallelism factors from only the reduction axis
and then \textit{lowers} synthesized programs to the full system hierarchy.
We \textit{prove} that \circled{3}, while largely reducing the search space,
is actually more expressive than \circled{2}.

\textit{Exploration with multiple reduction axes.}\,\,
The same observation applies for reduction over multiple axes. An example is given in
the second half of \Cref{fig:hierarchies}.
In this case, the reduction axes based synthesis hierarchy
is
{\small{$\begin{bmatrix} \makecmcolor{1}  & \makecmcolor{2} & \makecmcolor{3} & \makecmcolor{7} & \makecmcolor{8} & \makecmcolor{9}\end{bmatrix}$}}.
Note that
some parallelism factors are from the same hardware level:
$\makecmcolor{1}$ and $\makecmcolor{7}$,
$\makecmcolor{2}$ and $\makecmcolor{8}$,
and $\makecmcolor{3}$ and $\makecmcolor{9}$.
Since for switched networks,
splitting hardware hierarchies
does not bring benefits in most cases,
we can \textit{collapse}
parallelism factors of the same hardware hierarchies.
In this example, the final synthesis hierarchy is
{\small{$\begin{bmatrix} \makecmcolor{7}  & \makecmcolor{16} & \makecmcolor{27}\end{bmatrix}$}}.
\section{Program Synthesis}

We now present the program synthesis algorithm in \toolname.

\subsection{Parallelism Placement}
\label{sec:parallel-placement}

Parallelism placement partitions parallelism axes over the system hierarchy. With the novel notion of the \textit{parallelism matrix}
and its interpretation (\Cref{sec:overview:pp}), synthesizing
parallelism matrices is straightforward.
Consider\\
$\mathbf{H} = \begin{bmatrix} h_0 & \cdots & h_n \end{bmatrix}$ is the system hierarchy (e.g., {\small{$\begin{bmatrix}
    1 & 2 & 2& 4
\end{bmatrix}$}}),
$\mathbf{P} = \begin{bmatrix} p_0 & \cdots & p_m \end{bmatrix}$ is the parallelism axes (e.g., {\small{$\begin{bmatrix} 4 & 4 \end{bmatrix}$}}), \\
then a parallelism matrix is

\noindent
\begin{minipage}[t]{0.2\textwidth}
\vspace{-15pt}
\[
\begin{bmatrix} 
    x_{0,0} & x_{0,1} & \dots & x_{0, n} \\
    \vdots & \vdots & \ddots & \vdots \\
    x_{m,0} & x_{m,1} & \cdots  & x_{m, n} 
    \end{bmatrix}
\]
\end{minipage}
\begin{minipage}[t]{0.76\textwidth}
subject to:\\
$\mathlarger{\mathlarger{\prod}}_{i=0}^{m}x_{i,j} = h_j,\enskip j = 0, ..., n$ (1)
\\
$\mathlarger{\mathlarger{\prod}}_{j=0}^{n}x_{i,j} = p_i, \enskip i = 0, ..., m$ (2)
\end{minipage}

Equation (1)
requires the product of a column to be equivalent to the corresponding system hierarchy cardinality,
while Equation (2)
requires the product of a row to be equivalent to the corresponding parallelism axis.

\subsection{Collective Operations}
\label{sec:collective-operations}

This section defines the semantics of collective operations.
In this work we focus on the common ones:
AllReduce, ReduceScatter, AllGather, Reduce and Broadcast.

\paragraph{Notations}

We first define the notations.

\begin{tabular}{llll}\toprule
     $d$ & & &   device \\
    $s$ & $\in$ & $\bool^{k \times k }$ & device state \\
     $\G$ & $:=$  & $\overline{d_i : s_i}$  & state context \\
     $\op$ & $:=$ & $\allreduce \mid \reducescatter $  \\
           & $\mid$ & \multicolumn{2}{l}{$\allgather \mid \reduceno \mid \broadcastno $} \\
          \bottomrule
\end{tabular}

We use $d$ to denote a device,
whose state $s$ is represented as a boolean matrix of dimensions $k \times k$; $k$ being the number of devices. In particular we treat the data as being split in
$k$ chunks. The $i$th row of a state matrix represents the $i$th chunk.
$s[i][j] = 1$ means that device $j$ has contributed its original $i$th chunk
to the reduction result.
\Cref{fig:device:state}
gives an example.
A state context $\G$ maps devices to their states.

Note finally that Reduce and Broadcast typically take a root device to reduce to or broadcast from.
Since we focus on hierarchical systems,
we always use the first device in a reduction group as the root without loss of generality.


\begin{figure}
\begin{tikzpicture}[scale=0.3]
\draw[step=1cm,gray,very thin] (0,0) grid (4,4);
\foreach \x in {0,1}
  \foreach \y in {3}{
    \filldraw[mygrid] (\x,3-\x) rectangle (\x+1, 4-\x);
    \filldraw[mygrid] (\x+1,3-\x) rectangle (\x+2, 4-\x);
  }
\filldraw[mygrid] (2, 0) rectangle (3,1);
\filldraw[mygrid] (3, 0) rectangle (4,1);
\draw [decorate,decoration={brace,amplitude=5pt},xshift=-4pt,yshift=0pt]
(0,0) -- (0,4) node [black,midway,xshift=-1.2cm] 
{\footnotesize 3 data chunks};

\draw [decorate,decoration={brace,amplitude=5pt,mirror},xshift=4pt,yshift=0pt]
(4,3) -- (4,4) node [black,midway,xshift=2.1cm] 
{\footnotesize  reduced from device 0 and 1};

\draw [decorate,decoration={brace,amplitude=5pt,mirror},xshift=4pt,yshift=0pt]
(4,2) -- (4,3) node [black,midway,xshift=2.1cm] 
{\footnotesize  reduced from device 1 and 2};

\draw [decorate,decoration={brace,amplitude=5pt,mirror},xshift=4pt,yshift=0pt]
(4,0) -- (4,1) node [black,midway,xshift=2.1cm] 
{\footnotesize  reduced from device 2 and 3};
\end{tikzpicture}
\caption{A device state. Assume we have in total 4 devices (i.e., device 0,1,2 and 3),
so each device state is a $4 \times 4$ matrix.
$s[i][j]$ is colored if $s[i][j]=1$.
The device state has 3 non-empty rows, 
meaning that it has 3 data chunks.
Each data chunk describes where the data is reduced from.
For example, the first data chunk
is the reduction result between the original first data chunk of device 0 and 1.
}
\label{fig:device:state}
\end{figure}
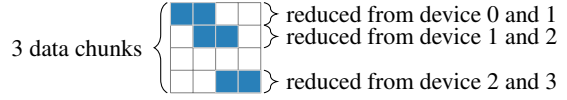

\begin{figure*}[t]

    

\centering

\begin{minipage}[t]{0.83\textwidth}

\renewcommand\MathparLineskip{\vspace{0pt}} 
    \begin{drulepar}[r]{$\hoarenoline{\G_1}{\op}{\G_2}$}{Reduction:
     from the pre-condition
     state $\G_1$, $\op$ yields to the post-condition state $\G_2$
    }
    \drule{allreduce}
    \drule{reducescatter}
    \vspace{3pt}
    \drule{allgather}
    \vspace{3pt}
    \drule{reduce}
    \vspace{8pt}
    
    \begin{minipage}{0.6\textwidth}
    \begin{tabular}{ll@{\hskip 40pt}ll} \toprule
     $\disjoint$ & \textit{disjoint} &
     $\mathsf{rows}$ & \textit{non-empty rows} \\
     $\union$  & \textit{addition} &
     $|\cdot|$ & \textit{length}\\
     \multicolumn{4}{l}{  $\mathsf{scatter}(s, \overline{i})$
     \textit{scatters non-empty rows in $s$ over devices $\overline{i}$}} \\
     \bottomrule
    \end{tabular}
    \end{minipage}
    \begin{minipage}{0.05\textwidth}
    \quad
    \end{minipage}
    \begin{minipage}{0.3\textwidth}
    \hspace{20pt}
    \vspace{5pt}
    \drule{broadcast}
    \end{minipage}
    \end{drulepar}
\end{minipage}
\begin{minipage}[t]{0.13\textwidth}
\renewcommand\ottaltinferrule[4]{ \inferrule*[narrower=1,lab=#1,#2] {#3} {#4}}
    \begin{tabular}[t]{p{0.7cm}!{\color{gray!40}\vrule}p{0.6cm} }
         before & after\\[15pt]
        \begin{tikzpicture}[scale=0.14]
        \draw[step=1cm,gray,very thin] (0,0) grid (4,4);
        \filldraw[mygrid] (0,3) rectangle (1,4);
        \filldraw[mygrid] (2,3) rectangle (3,4);
        \filldraw[mygrid] (2,1) rectangle (3,2);
        \end{tikzpicture}
        \vspace{4pt}
        \begin{tikzpicture}[scale=0.14]
        \draw[step=1cm,gray,very thin] (0,0) grid (4,4);
        \filldraw[mygrid] (1,1) rectangle (2,2);
        \filldraw[mygrid] (3,1) rectangle (4,2);
        \filldraw[mygrid] (1,3) rectangle (2,4);
        \filldraw[mygrid] (3,3) rectangle (4,4);
        \end{tikzpicture}
        &
        \begin{tikzpicture}[scale=0.14]
        \draw[step=1cm,gray,very thin] (0,0) grid (4,4);
        \foreach \x in {1,2,3}
          \foreach \y in {1,3}
            \filldraw[mygrid] (\x,\y) rectangle (\x+1,\y+1);
        \filldraw[mygrid] (0,3) rectangle (1,4);
        \end{tikzpicture}
        \newline
        \begin{tikzpicture}[scale=0.14]
        \draw[step=1cm,gray,very thin] (0,0) grid (4,4);
        \foreach \x in {1,2,3}
          \foreach \y in {1,3}
            \filldraw[mygrid] (\x,\y) rectangle (\x+1,\y+1);
        \filldraw[mygrid] (0,3) rectangle (1,4);
        \end{tikzpicture}
         \\[15pt]
        \begin{tikzpicture}[scale=0.14]
        \draw[step=1cm,gray,very thin] (0,0) grid (4,4);
        \filldraw[mygrid] (0,3) rectangle (1,4);
        \filldraw[mygrid] (2,3) rectangle (3,4);
        \filldraw[mygrid] (2,1) rectangle (3,2);
        \end{tikzpicture}
        \newline
        \begin{tikzpicture}[scale=0.14]
        \draw[step=1cm,gray,very thin] (0,0) grid (4,4);
        \filldraw[mygrid] (1,1) rectangle (2,2);
        \filldraw[mygrid] (3,1) rectangle (4,2);
        \filldraw[mygrid] (1,3) rectangle (2,4);
        \filldraw[mygrid] (3,3) rectangle (4,4);
        \end{tikzpicture}
        &
        \begin{tikzpicture}[scale=0.14]
        \draw[step=1cm,gray,very thin] (0,0) grid (4,4);
        \foreach \x in {0,1,2,3}
          \foreach \y in {3}
            \filldraw[mygrid] (\x,\y) rectangle (\x+1,\y+1);
        \end{tikzpicture}
        \newline
        \begin{tikzpicture}[scale=0.14]
        \draw[step=1cm,gray,very thin] (0,0) grid (4,4);
        \foreach \x in {1,2,3}
          \foreach \y in {1}
            \filldraw[mygrid] (\x,\y) rectangle (\x+1,\y+1);
        \end{tikzpicture}
         \\[20pt]
        \begin{tikzpicture}[scale=0.14]
        \draw[step=1cm,gray,very thin] (0,0) grid (4,4);
        \foreach \x in {0,2}
          \foreach \y in {1, 3}
            \filldraw[mygrid] (\x,\y) rectangle (\x+1,\y+1);
        \end{tikzpicture}
        \newline
        \begin{tikzpicture}[scale=0.14]
        \draw[step=1cm,gray,very thin] (0,0) grid (4,4);
        \foreach \x in {1,3}
          \foreach \y in {0, 2}
            \filldraw[mygrid] (\x,\y) rectangle (\x+1,\y+1);
        \end{tikzpicture}
        &
        \begin{tikzpicture}[scale=0.14]
        \draw[step=1cm,gray,very thin] (0,0) grid (4,4);
        \foreach \x in {0,2}
          \foreach \y in {1, 3}
            \filldraw[mygrid] (\x,\y) rectangle (\x+1,\y+1);
        \foreach \x in {1,3}
          \foreach \y in {0, 2}
            \filldraw[mygrid] (\x,\y) rectangle (\x+1,\y+1);
        \end{tikzpicture}
        \newline
        \begin{tikzpicture}[scale=0.14]
        \draw[step=1cm,gray,very thin] (0,0) grid (4,4);
        \foreach \x in {0,2}
          \foreach \y in {1, 3}
            \filldraw[mygrid] (\x,\y) rectangle (\x+1,\y+1);
        \foreach \x in {1,3}
          \foreach \y in {0, 2}
            \filldraw[mygrid] (\x,\y) rectangle (\x+1,\y+1);
        \end{tikzpicture}
         \\[20pt]
        \begin{tikzpicture}[scale=0.14]
        \draw[step=1cm,gray,very thin] (0,0) grid (4,4);
        \filldraw[mygrid] (0,3) rectangle (1,4);
        \filldraw[mygrid] (2,3) rectangle (3,4);
        \filldraw[mygrid] (2,1) rectangle (3,2);
        \end{tikzpicture}
        \vspace{4pt}
        \begin{tikzpicture}[scale=0.14]
        \draw[step=1cm,gray,very thin] (0,0) grid (4,4);
        \filldraw[mygrid] (1,1) rectangle (2,2);
        \filldraw[mygrid] (3,1) rectangle (4,2);
        \filldraw[mygrid] (1,3) rectangle (2,4);
        \filldraw[mygrid] (3,3) rectangle (4,4);
        \end{tikzpicture}
        &
        \begin{tikzpicture}[scale=0.14]
        \draw[step=1cm,gray,very thin] (0,0) grid (4,4);
        \foreach \x in {1,2,3}
          \foreach \y in {1,3}
            \filldraw[mygrid] (\x,\y) rectangle (\x+1,\y+1);
        \filldraw[mygrid] (0,3) rectangle (1,4);
        \end{tikzpicture}
        \newline
        \begin{tikzpicture}[scale=0.14]
        \draw[step=1cm,gray,very thin] (0,0) grid (4,4);
        \end{tikzpicture}
        \\[30pt]
        \begin{tikzpicture}[scale=0.14]
        \draw[step=1cm,gray,very thin] (0,0) grid (4,4);
        \foreach \x in {1,2,3}
          \foreach \y in {1,3}
            \filldraw[mygrid] (\x,\y) rectangle (\x+1,\y+1);
        \filldraw[mygrid] (0,3) rectangle (1,4);
        \end{tikzpicture}
        \newline
        \begin{tikzpicture}[scale=0.14]
        \draw[step=1cm,gray,very thin] (0,0) grid (4,4);
        \filldraw[mygrid] (1,1) rectangle (2,2);
        \filldraw[mygrid] (3,1) rectangle (4,2);
        \filldraw[mygrid] (1,3) rectangle (2,4);
        \filldraw[mygrid] (3,3) rectangle (4,4);
        \end{tikzpicture}
        &
        \begin{tikzpicture}[scale=0.14]
        \draw[step=1cm,gray,very thin] (0,0) grid (4,4);
        \foreach \x in {1,2,3}
          \foreach \y in {1,3}
            \filldraw[mygrid] (\x,\y) rectangle (\x+1,\y+1);
        \filldraw[mygrid] (0,3) rectangle (1,4);
        \end{tikzpicture}
        \newline
        \begin{tikzpicture}[scale=0.14]
        \draw[step=1cm,gray,very thin] (0,0) grid (4,4);
        \foreach \x in {1,2,3}
          \foreach \y in {1,3}
            \filldraw[mygrid] (\x,\y) rectangle (\x+1,\y+1);
        \filldraw[mygrid] (0,3) rectangle (1,4);
        \end{tikzpicture}
    \end{tabular}
\end{minipage}
    \caption{Semantics of collective operations.
    with the right presents examples of each operation. For those
    examples, we have in total 4 devices .e., device 0,1,2, and 3),
    so each device state is a $4 \times 4$ matrix.
    We assume
    the reduction happens between only device 0 and 1.
    The pre-condition states of device 0 (top) and 1 (bottom) are on the left,
    and after a step of reduction, their states turn into the
    post-condition states on the right.}
    \label{fig:op:semantics}
\end{figure*}


\paragraph{Semantics}

\Cref{fig:op:semantics} defines the semantics of collective operations,
which is closely based on Hoare rules \cite{hoare}.
Each reduction takes the form of a Hoare triple $\hoarenoline{\G_1}{\op}{\G_2}$, which means that
\textit{from the pre-condition
state $\G_1$, a step of reduction $\op$ yields to the post-condition state $\G_2$}.
Explanations
of auxiliary functions are given in the figure.
To better illustrate the semantics, the right of
\Cref{fig:op:semantics} provides examples of each collective operation.

At a high level, these rules capture
the constraints for a reduction step to be semantically correct.
\Rref{r-AllReduce} first checks that
the data contained in each device (denoted as $\mathsf{rows}$ representing the \textit{non-empty rows})
should have the same data chunks. Moreover, columns in any specific chunk
should be disjoint. Both constraints are essential for the reduction result to be valid:
we should not reduce 
data from different chunks or reduce the same data twice
(as discussed in \Cref{sec:overview:collective:operations}).
Finally, we generate the resulting state $\union \overline{s_i}$ for each device by adding up all matrices.
\Rref{s-ReduceScatter} and \rref*{s-Reduce} are similar to \rref{s-AllReduce}, except that \rref*{s-ReduceScatter} scatters the reduction result over devices, where
$\mathsf{scatter}$ raises an error
if the number of data chunks in $s$ is not divisible by the number of devices;
and \rref*{s-reduce} puts the result only in the first device and clears up the rest of the devices. 
\Rref{s-AllGather} simply needs all data rows to be disjoint.
\Rref{s-Broadcast} overrides the data of every device with the data from the first one.
As an optimization, the rule enforces \textit{information increase}, i.e.,
the data to be broadcasted must be as informative as data in other devices
and more informative than at least one other device.

\subsection{Reduction Programs}
\label{sec:reduction-programs}

We now turn to our reduction language which is built on top of the formalism of collective operations.

\begin{tabular}{p{1.3cm}lll}\toprule
    $\program$ & $\in$ & $[\reduction]$\\
    $\reduction$ & $\in $ & ${\slice} \times {\form} \times {\op}$ \\
    $\slice$ & $:=$ & $e$ \\
    $\form$ & $:=$ & $\insidegroup \mid \parallel{e} \mid \master{e}$  \\ \bottomrule
\end{tabular}

A reduction strategy is represented as a $\program$, which is essentially a list of reduction instructions.
A $\reduction$ instruction consists of a $\slice$, a $\form$, and a collective operation $\op$.
We use $e$ to represent a level in the synthesis hierarchy.
The $\slice$ chooses a level.
The $\form$ has three patterns: $\insidegroup$,
$\parallel{e}$, and $\master{e}$.
Inside a $\reduction$,
the $e$ carried in the form must be an ancestor of the one carried in the slice.
The slice and the form together decide the device groups
that will perform the operation $\op$.

It turns out that $\slice$ and $\form$ are quite expressive and
can encode many common hierarchical communication patterns.
\Cref{fig:reduction:example:eg} demonstrates several examples using the system hierarchy
in \Cref{fig:overview:1}.
Specifically,
a $\slice$ divides devices into \textit{reduction groups}, and $\form$ decides the reduction form
happening for the reduction groups.
For example, consider that the slice is CPU,
then we get reduction groups within each CPU, i.e., 
{\scriptsize{$\mathsf{\group{A_0,A_1,A_2,A3}, \group{B_0,B_1,B_2,B_3}, \group{C_0,C_1,C_2,C_3}, \group{D_0,D_1,D_2,D_3}}$}}.
Now, if the form is $\insidegroup$, then we perform reduction within each reduction group.
If the form is $\parallel{e}$, we perform reduction over the first device in each group, the second device in each group, etc,
if they connect to the same $e$.
Thus, $\parallel{\text{server}}$ generates {\scriptsize{$\group{\mathsf{A_0,B_0}}$}},
{\scriptsize{$\group{\mathsf{A_1,B_1}}$}}, etc., whereas $\parallel{\text{rack}}$ 
generates {\scriptsize{$\mathsf{\group{A_0,B_0,C_0,D_0}}$}} etc.
$\masterno$ generates the device groups in the same way as $\parallelno$, but only reduces over the first
device group.

Note that \Cref{fig:reduction:example:eg} presents device groups for
reduction over the system hierarchy
[(rack, 1), (server, 2), (CPU, 2), (GPU, 4)].
As we discussed in \Cref{sec:overview:hierarchy},
reduction over specific parallelism axes
will use the synthesis hierarchy formed by parallelism factors,
and we will get reduction groups for that particular reduction axis like {\scriptsize{$\group{\mathsf{A_0,A_1}},
\group{\mathsf{A2,A3}}$}} etc.

\begin{table}[t]
    \centering
    \begin{scriptsize}
    \setlength\tabcolsep{4.5pt}
    \begin{tabular}{lll} \toprule
        $\slice$ & $\form$ & $\mathsf{groups}(\slice,\form)$  \\ \hline  \rule{0pt}{1.2\normalbaselineskip}
        CPU & $\insidegroup$ & 
        \scriptsize{$\mathsf{\group{A_0,A_1,A_2,A_3}, \group{B_0,B_1,B_2,B_3}, }$} \\
        & &  \scriptsize{$\mathsf{\group{C_0,C_1,C_2,C_3}, \group{D_0,D_1,D_2,D_3}} $} \\
         & $\parallel{\text{server}}$ & 
        \scriptsize{$\mathsf{\group{A_0,B_0},\group{A_1, B_1}, \group{A_2,B_2}, \group{A_3,B_3} }$} \\ 
        & &  \scriptsize{$\mathsf{\group{C_0,D_0}, \group{C_1, D_1}, \group{C_2,D_2}, \group{C_3,D_3}} $} \\
         & $\parallel{\text{rack}}$ & 
        \scriptsize{$\mathsf{\group{A_0,B_0,C_0,D_0},\group{A_1,B_1,C_1,D_1}, }$} \\ 
        & &  \scriptsize{$\mathsf{\group{A_2, B_2, C_2,D_2}, \group{A_3, B_3, C_3,D_3}} $} \\
         & $\master{\text{rack}}$ & 
        \scriptsize{$\mathsf{\group{A_0,B_0,C_0,D_0}}$} \\ 
        server & $\insidegroup$ & 
        \scriptsize{$\mathsf{\group{A_0,A_1,A_2, A_3, B_0,B_1,B_2,B_3}, }$} \\ 
        & &  \scriptsize{$ \mathsf{\group{C_0,C_1,C_2, C_3,D_0,D_1,D_2,D_3}} $} \\
         & $\parallel{\text{rack}}$ & 
        \scriptsize{$\mathsf{\group{A_0,C_0},\group{A_1, C_1}, \group{A_2,C_2}, \group{A_3,C_3}}$} \\ 
        & &  \scriptsize{$\mathsf{\group{B_0,D_0}, \group{B_1, D_1}, \group{B_2,D_2}, \group{B_3,D_3}}$} \\
        rack & $\insidegroup$ & 
        \scriptsize{$\mathsf{\{A_0,A_1,A_2,A_3,B_0,B_1,B_2,B_3,}$} \\
        & & \scriptsize{$\mathsf{\enskip C_0,C_1,C_2,C_3,D_0,D_1,D_2,D_3\}}$} \\
        \bottomrule
    \end{tabular}
    \end{scriptsize}
    \caption{Hierarchical communication patterns for \Cref{fig:overview:1}.}
    \label{fig:reduction:example:eg}
\end{table}

Supposing $\slice$ and $\form$ derive the device groups $\overline{\G_i}$,
which are disjoint by construction,
we define the semantics of a $\reduction$ instruction as:
\begin{mathpar}\small
\renewcommand\ottaltinferrule[4]{  \inferrule*[narrower=1] {#3} {#4} }
\drule{reduction}
\end{mathpar}
where each device group participating in the reduction gets the device states updated
according to the semantics of collective operations,
and devices not participating in the reduction have their states unchanged.
A reduction program then iteratively applies each reduction:
\begin{mathpar}\small
\renewcommand\ottaltinferrule[4]{  \inferrule*[narrower=1] {#3} {#4} }
\drule{program}
\end{mathpar}

\subsection{Synthesis Hierarchy}
\label{sec:system-hierarchy}

In \Cref{sec:overview:hierarchy}, we have proposed and compared different synthesis
hierarchies for synthesizing reduction programs:

(a) System hierarchy ({\small{$\begin{bmatrix}1&2&2&4\end{bmatrix}$}}) \\
(b) Column-based parallelism factors ({\small{$\begin{bmatrix} 1 & \makecmcolor{1} & 1 & \makecmcolor{2} & 2 & \makecmcolor{1} & 2 & \makecmcolor{2}\end{bmatrix}$}}) \\
(c) Row-based parallelism factors ({\small{$\begin{bmatrix}1 & 1 & 2 & 2 & \makecmcolor{1} & \makecmcolor{2} & \makecmcolor{1} & \makecmcolor{2} \end{bmatrix}$}})\\
(d) Reduction axis parallelism factors ({\small{$\begin{bmatrix}\makecmcolor{1} & \makecmcolor{2} & \makecmcolor{1} & \makecmcolor{2} \end{bmatrix}$}})

\toolname uses (d).
Here, we formally prove the theorem that justifies our choice.
First, every reduction instruction essentially decides the device groups $\overline{\G}$ and the
operation $\op$. Therefore, a program can be lowered to a sequence $(\overline{\G_1},\op_1)$, $(\overline{\G_2},\op_2)$, ..., $(\overline{\G_n}, \op_n)$.
Since (d) includes only the reduction axis,
lowering for (d)
applies the generated grouping patterns
to non-reduction axes when forming device groups.
\begin{definition}[Expressive power of synthesis hierarchy]
A synthesis hierarchy is more expressive than ($\geq$) another,
if every valid lowered program $\mathcal{L}$ synthesized using the latter can be synthesized using the former.
\end{definition}
\begin{restatable}{theorem}{thmexpressive}
\label{thm:expressive}
(d) $\geq$ (c) $\geq$ (b) $\geq$ (a).
\end{restatable}
\proof
For space reasons, we show one example that illustrates the key proof strategy,
and we refer the reader to the appendix for the full proof.
Consider proving (c) $\geq$ (b), where (b) synthesizes a valid reduction step
$(e_2, \parallel{e_1}, \op)$.
For the reduction to be valid, every
reduction group must be partitioned only over the reduction axis.
Therefore, all non-reduction parallelism factors column-wisely
between $e_1$ (exclusive) and $e_2$ (inclusive)
can only be $1$.
An example is given below on the left.\\
Now we construct a reduction instruction for (c) that expresses the same reduction.
Suppose the reduction step in (b) covers parallelism factors $e_i,...,e_j$ on the reduction axis.
Let $e_1'$ be the level corresponding to the parallelism factor right
before $e_i$ row-wisely,
and let $e_2'$ be $e_j$. Then $(e_2', \parallel{e_1'}, \op)$ is a desired reduction instruction.\\
\tikzset{%
    myarrow/.style = {-Stealth, shorten >=5pt}
} 
\newcommand{\mypoint}[2]{\tikz[remember picture]{\node[inner sep=0, text centered] (#1)  {$#2$};}}
\begin{minipage}{0.07\textwidth}
\quad
\end{minipage}
\begin{minipage}{0.18\textwidth}
\begin{small}
\begin{gather*}
\begin{bmatrix} 
    x_{0,0}  & 1 & 1 & x_{0,3} \\
    \mypoint{e1}{\bm{{x_{1,0}}}} & 1 & 1 & x_{1,3} \\
    \mypoint{e3}{\makecmcolor{x_{2,0}}} & \makecmcolor{x_{2,1}} & \makecmcolor{x_{2,2}} & \makecmcolor{x_{2,3}} \\
    1 & 1 & \mypoint{e2}{\bm{1}} & x_{3,3} \\
\end{bmatrix}
\end{gather*}
 \begin{tikzpicture}[remember picture, overlay]
      \def\arraystretch{0.2}
      \node[left=15pt of e1](textofhere1){$e_1$};
      \draw[myarrow] (textofhere1) -- (e1);
      \node[right=25pt of e2](textofhere2){$e_2$};
      \draw[myarrow] (textofhere2) -- (e2);
      \node[left=11pt of e3](textofhere3){\begin{tabular}{r}reduction\\ axis\end{tabular}};
      \draw[myarrow,line width=2pt,color=cmcolor] (textofhere3) -- (e3);
    \end{tikzpicture}
\end{small}
\end{minipage}
\begin{minipage}{0.21\textwidth}
\begin{small}
\begin{gather*}
\begin{bmatrix} 
    x_{0,0}  & 1 & 1 & x_{0,3} \\
    x_{1,0} & 1 & 1 & \mypoint{e1}{\bm{{x_{1,3}}}} \\
    \makecmcolor{{x_{2,0}}} & \makecmcolor{x_{2,1}} & \mypoint{e2}{\makecmcolor{\bm{x_{2,2}}}} & \makecmcolor{x_{2,3}} \\
    1 & 1 & 1 & x_{3,3} \\
\end{bmatrix}
\end{gather*}
 \begin{tikzpicture}[remember picture, overlay]
      \node[right=15pt of e1](textofhere1){$e_1'$};
      \draw[myarrow] (textofhere1) -- (e1);
      \node[right=25pt of e2](textofhere2){$e_2'$};
      \draw[myarrow] (textofhere2) -- (e2);
    \end{tikzpicture}
\end{small}
\end{minipage}
\qed

\subsection{Program Synthesis for Reduction Programs}

So far, we have given the constraint for generating parallelism matrices (\Cref{sec:parallel-placement}) and how we can obtain a synthesis hierarchy
from a parallelism matrix  (\Cref{sec:system-hierarchy}). The last missing piece is how to synthesize reduction programs.

To formalize the synthesis problem, we need an initial pre-condition state as the beginning
state and a post-condition state as the final desired state.
Initially, every device only holds its own data, and therefore
device $i$ has 1 in the $i$th column, and 0 in any other position.
In the final desired state, a device should have 1 in all columns
corresponding to devices in its reduction group, and 0 in any other position.
An extra indirection is caused from using the reduction axis parallelism factors as the synthesis hierarchy (\Cref{sec:system-hierarchy}), which only includes part of the system,
and then lowers the program to the full system.
Therefore, our goal is to synthesize a $\program$,
whose lowering $\mathcal{L}$ subjects to:
\[
\scriptsize{
\left\{ \overline{d_i:\begin{array}{c}
\begin{matrix}
 &  & \enskip i & & &
\end{matrix} \\
\begin{bmatrix} 
    0 & \dots & 1 & \dots &   0 \\
    \vdots & \ddots & \vdots & \ddots & \vdots \\
    0 & \cdots & 1  & \dots & 0 
    \end{bmatrix}
\end{array} }
\right\}
{\mathcal{L}}
\left\{
\overline{
d_i: \begin{array}{c}
\begin{matrix}
\enskip  & i & \,\,\,\,  &  \,\,\,\, & \,\,\,\,  & \,\,\,\, & \  & \overline{j} & \quad
\end{matrix} \\
\begin{bmatrix} 
    0 & \dots & 1 & \dots & 0 & \dots & 1 & \dots & 0  \\
    \vdots & \ddots & \vdots & \ddots & \vdots & \ddots & \vdots & \ddots & \vdots \\
    0 & \cdots & 1  & \dots & 0 & \cdots & 1 & \cdots & 0 
    \end{bmatrix}
\end{array}   
}
\right\}
}
\]
supposing $d_i$ reduces with devices $\overline{j}$.

Given the syntax and the semantics of reduction programs,
we use \textit{syntax-guided program synthesis}~\cite{alur2013syntax} to synthesize programs
in increasing order of program size.

\begin{figure}[t]\footnotesize
    \centering
    \begin{subfigure}{\linewidth}
    \includegraphics[width=\linewidth]{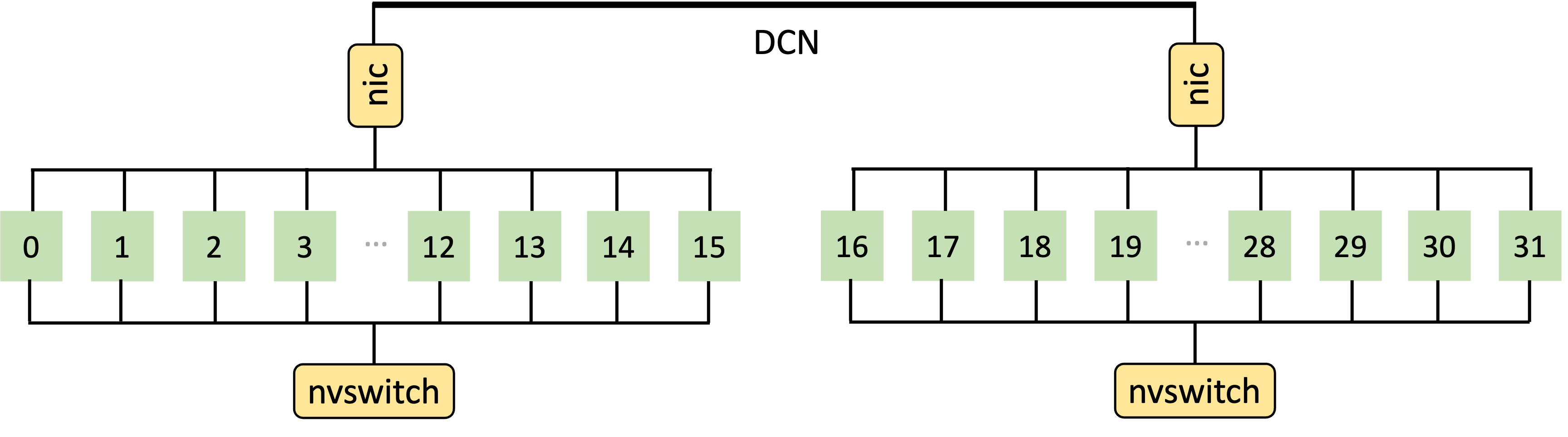}
    \caption{2 nodes, each with 16 A100 GPUs sharing one NVSwitch and one NIC,
    and all NICs are connected in a data center}
    \label{fig:experiments:a}
    \end{subfigure}
    
    \begin{subfigure}{\linewidth}
    \includegraphics[width=\linewidth]{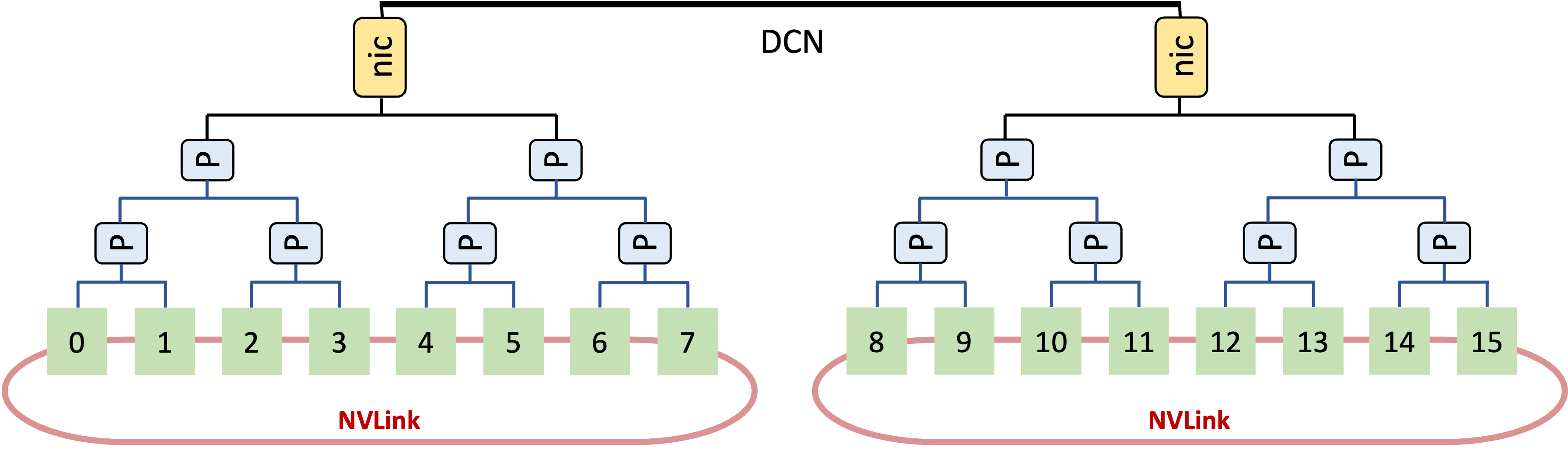}
    \caption{2 nodes, each with 8 V100 GPUs forming a ring via NVLink and connected via PCIe switches. Each node consists of two
    CPUs (each owning 4 GPUs) with one NIC
    to the DCN.
    A shared NIC connecting the two CPUs is a modeling simplification --
    in reality cross-domain communication is through shared memory.}
    \label{fig:experiments:b}
    \end{subfigure}
    \caption{System topology models for 2 nodes. For the experiments, we run on both 2 and 4 nodes.}
    \label{fig:experiment:system}
\end{figure}

\begin{table}[t]
\footnotesize
\centering
 \setlength{\tabcolsep}{3pt}
\begin{tabular}{cM{1.5cm}M{1.9cm}rrrr} 
     \hline
     \rowcolor{headrowcolor}
     &Parallelism axes & Parallelism matrix 
     & \multicolumn{2}{M{1.9cm}}{Reduction on the 0th axis\newline Ring $\,\,\,\,$  Tree}  
     & \multicolumn{2}{M{1.9cm}}{Reduction on the 1st axis\newline Ring $\,\,\,\,$  Tree}  
     \\ \hline
     \multicolumn{5}{l}{4 nodes, each with 16 A100} \\ \hline
     \rowcolor{rowcolor}
     A1 & \axes{2 & 32}
     & \axes{\axesinside{1 & 2} & \axesinside{4 & 8} } & 0.12 & 0.17 & 8.74   & 9.89 \\
     A2 && \axes{ \axesinside{2 & 1}& \axesinside{2 & 16} }& 37.16 & 36.94 & 4.81  & 3.41  \\[0.5pt] \hline
     \rowcolor{rowcolor}
     B1 & \axes{4 & 16}
     & \axes{\axesinside{1 & 4} & \axesinside{4 & 4} } & 0.15 & 0.20 & 17.70   & 19.03 \\
     B2 & & \axes{\axesinside{2 & 2} & \axesinside{2 & 8} } & 28.77  & 19.81 & 8.39  & 4.99  \\ 
     B3 & & \axes{ \axesinside{4 & 1}& \axesinside{1 & 16} }& 56.13  & 89.70 & 0.18   & 0.22  \\[0.5pt] \hline
     \rowcolor{rowcolor}
     C1 & \axes{8 & 8} 
     & \axes{\axesinside{1 & 8} & \axesinside{4 & 2} } & 0.17  & 0.21 & 33.92  & 41.06  \\
     C2 && \axes{\axesinside{2 & 4} & \axesinside{2 & 4} } & 16.52 & 9.18 & 15.68  & 9.43 \\
     C3 && \axes{\axesinside{4 & 2} & \axesinside{1 & 8} } & 34.05 & 41.23 & 0.17   & 0.21  \\[0.5pt] \hline
     \multicolumn{5}{l}{4 nodes, each with 8 V100} \\ \hline
     \rowcolor{rowcolor}
     E1 & \axes{8 & 4}
     & \axes{\axesinside{1 & 8} & \axesinside{4 & 1} }      & 0.28  & 0.39  & 21.74  & 30.42 \\
     E2 & & \axes{\axesinside{2 & 4} & \axesinside{2 & 2} } & 14.25 & 15.48 & 10.98  & 7.34  \\ 
     E3 & & \axes{ \axesinside{4 & 2}& \axesinside{1 & 4} } & 14.84 & 19.90 & 2.96   & 0.43  \\
     \hline
\end{tabular}
\caption{Reduction time in seconds of running $\allreduce$.
 }
 \label{table:experiment}
\end{table}

\section{Experiments}
\label{sec:evaluation}

We implement \toolname to synthesize parallelism matrices and reduction programs,
and lower the programs into sequences of XLA collective operations, which in turn result in sequences of NCCL calls on the XLA GPU backend.
\ningning{give nccl version/cuda version/etc?}
We measure the execution time of the compiled programs.
The experiments aim to answer the following research questions:

\textbf{RQ1}$\quad$ What is the impact of parallelism placement on reduction algorithms?

\textbf{RQ2}$\quad$ Are our various techniques for taming the search space effective
so that we can quickly enumerate a wide variety of reduction programs?

\textbf{RQ3}$\quad$ Given a parallelism placement, can we find reduction strategies that outperform the default implementation (i.e., $\allreduce$), and if so what is their form?

The experiments ran on two different GPU system configurations
available on Google Cloud Platform~\cite{gcp-gpus}
(see \Cref{fig:experiment:system}):
(i) NVIDIA A100,
where each node consists of 16 GPUs sharing one NVSwitch and one NIC connecting to the data center network; and (ii) NVIDIA Tesla V100, where each node consists of 8 GPUs
forming a {\em ring} via NVLink; each pair of GPUs are connected via PCIe switches,
and each of
the two CPUs of the node has 4 GPUS in its PCIe domain.
We experiment with both NCCL ring reduction and tree reduction~\cite{sanders2009two}, set by NCCL\_ALGO.

\begin{table*}[t]
\footnotesize
\centering
\begin{tabular}{cM{0.8cm}M{1.6cm}M{1.2cm}M{2.2cm}lM{1.6cm}M{1.4cm}c} \hline
 \rowcolor{headrowcolor}
 & NCCL algo 
 & Parallelism axes
 & Synthesis time (s)
 & Programs outperforming $\allreduce$ / total programs
 & Parallelism matrix 
 & $\allreduce$ (bold if the optimal $\allreduce$)
 & Optimal (bold if overall optimal)
 & Speedup\\ \hline \multicolumn{8}{l}{2 nodes, each with 16 A100} \\ \hline
     \rowcolor{rowcolor}
     F1 &
     Ring &
     \axes{8&4}  &
     0.03 &
     14/47 & \axes{\axesinside{1 & 8} \axesinside{2 & 2} } & \textbf{0.17}  & \textbf{0.17} & 1$\times$ 
     \\
     F2 &&&&& \axes{\axesinside{2&4} \axesinside{1&4}} & 16.84  & 9.19 & 1.83$\times$ \\
     \hline
     \multicolumn{8}{l}{4 nodes, each with 16 A100} \\ \hline
     \rowcolor{rowcolor}
     G1& Tree &
     \axes{{4} & 16} &
     0.04 &
     10/53   & \axes{\axesinside{1 & 4} \axesinside{4 & 4}}& \textbf{0.20} & \textbf{0.17} & 1.17$\times$ \\
     G2 &&&&& \axes{\axesinside{4 & 1} \axesinside{1 & 16}}& 89.70 & 56.13 & 1.60$\times$ \\
     \rowcolor{rowcolor}
     H1 & Ring &
     \axes{{16} & 2 & {2}}  &
     0.97 &
     25/235    & \axes{\axesinside{1&16} \axesinside{2&1} \axesinside{2&1}} & \textbf{4.79} & 4.63  & 1.03$\times$ 
     \\
     H2&&&&           & \axes{\axesinside{2&8} \axesinside{2&1} \axesinside{1&2}} & 4.91 & \textbf{3.10} & 1.58$\times$ 
     \\
     \rowcolor{rowcolor}
     I1 & 
     Ring & 
     \axes{{2} & 2 & {16}} & 
     0.93 & 
      29/235 & \axes{\axesinside{2 & 1} \axesinside{2 & 1} \axesinside{1 & 16} } &
     \textbf{4.82} & \textbf{2.99} & 1.61$\times$ 
     \\
     I2 &&&&&
     \axes{\axesinside{1 & 2} \axesinside{2 & 1} \axesinside{2 & 8} } &
     5.28 & 4.77 & 1.11$\times$ 
     \\
     \rowcolor{rowcolor}
     J1 &
     Tree &
     \axes{{64}} & 
     1.16 &
     5/47 & \axes{\axesinside{4 & 16}} & \textbf{5.75} & \textbf{4.74} & 1.21$\times$ 
     \\
     \hline
     \multicolumn{8}{l}{4 nodes, each with 8 V100} 
     \\ \hline
     \rowcolor{rowcolor}
     K1 & 
     Ring & 
     \axes{{8}&2&{2}}  &
     0.24 &
     17/188  & \axes{\axesinside{2 & 4} \axesinside{2 & 1} \axesinside{1 & 2}} &  4.80  & \textbf{2.35}  & 2.04$\times$  
     \\
     K2&&
     &   &  & \axes{\axesinside{1 & 8} \axesinside{2 & 1} \axesinside{2 & 1}} &  \textbf{4.40}  & 4.40  & 1$\times$  
     \\
     \rowcolor{rowcolor}
     L1 & 
     Ring & 
     \axes{{32}}  &
     0.06 &
     11/47   & \axes{\axesinside{4 & 8}} & \textbf{4.83}  & \textbf{3.45} & 1.4$\times$  \\
     \hline
\end{tabular}
\caption{Reduction time in seconds for running $\allreduce$ and the synthesized optimal reduction strategy (reduction on the 0th axis for parallelism axes of size 1 and 2, and
on the 0th and 2rd axes for parallelism axes of size 3).
 }
 \label{table:experiment:2}
\end{table*}

We run experiments with 2 and 4 nodes. For A100, the system hierarchy
is {\small{$\begin{bmatrix} 2 & 16 \end{bmatrix}$}} or  {\small{$\begin{bmatrix} 4 & 16 \end{bmatrix}$}}.
For V100, since the NVLink ring connects all 8 GPUs,
and the NVLink ring has much higher bandwidth than PCIe bridges,
we put 8 GPUs inside one layer, and so with 2 or 4 nodes, the system hierarchy is 
{\small{$\begin{bmatrix} 2 & 8 \end{bmatrix}$}} or
{\small{$\begin{bmatrix} 4 & 8 \end{bmatrix}$}}, respectively.
Each GPU carries a large amount of data ($(2^{29} \times$ nodes) of float32) to reduce the impact of latency, and each program runs 10 times to reduce the impact of network noise.

For each system, we synthesize parallelism mappings and reduction programs
for (1) a single parallelism axis;
(2) all combinations of two parallelism axes, with reduction on one of the axes; and
(3) three parallelism axes, with reduction on the first and the third axes.
We can easily scale to more axes, though up to three axes are
quite common in practical settings,
and many observations can already be illustrated.

Next, we discuss the results and insights from the experiments. For space reasons,
we present only representative cases,
and we put the full experiment results in the appendix.

\subsection{Synthesizing Parallelism Placement}

\textbf{Result 1} (\textbf{RQ 1}):
\textit{
The performance of $\allreduce$
differs significantly
among parallelism matrices,
up to 448.5$\times$}.

The experiment results are given in \Cref{table:experiment}.
For a particular parallelism axis (e.g., A),
we compare the reduction time for difference parallelism matrices (e.g., A1 and A2)
with each NCCL algorithm and the reduction axis. Notably,
for reducion on the 0th axis and with the Tree algorithm,
B3 (89.70s) is slower than B1 (0.20s) by 448.5$\times$. 

The difference is due to the fact that different parallelism matrices
lead to different data placement. In B1, 
the first row of the the matrix
(\axes{1 & 4}) means that devices to be reduced are inside a single node,
where the local NVSwitch can perform the reduction efficiently.
For B3, the first row (\axes{4 & 1}) puts reduction groups across nodes, going 
through the slow data-center network.
However, B3 can still be useful {\em for a diffferent reduction}:
since it puts the 1st reduction axis inside a single node,
for a reduction on the 1st axis,
B3 (0.22s on Tree) is 86.5$\times$ faster than B1 (19.03s).
In practice, models with multiple parallelism forms (e.g., \citet{shoeybi2020megatronlm}) involve reductions across both axes, and the selection of a mapping should take all of them into account.

\subsection{Synthesizing Reduction Programs}

Now we turn to the reduction programs synthesized for each parallelism matrix.
\Cref{table:experiment:2} presents experiment results.

\textbf{Result 2} (\textbf{RQ 2}): \textit{Our pruning techniques
are effective for the synthesizer to achieve fast synthesis time.}

With our formalism, a program cannot be arbitrarily large,
since our carefully crafted semantics
of collective operations enforces a form of information increase for every operation.
In our experiments, we set 5 as the program size limit for the synthesizer,
which turns out to be sufficient to generate interesting reduction patterns.
With this setup,
the longest synthesis time is under 2 seconds (for up to 235 programs).
Increasing the size limit makes
the synthesis \textit{slightly} slower, but, for most cases, does not generate new programs.

\textbf{Result 3} (\textbf{RQ 3}): \textit{
If the reduction axes can be put within one node,
then a single step $\allreduce$
inside that node is the most performant reduction due to fast local bandwidth}.

We observe this result from the difference between F1 and F2.
F1 assigns the reduction axis to the GPU level, and thus
$\allreduce$ is the most performant reduction,
outperforming F2, which requires cross-node reduction, by 99.06$\times$.

\textbf{Result 4} (\textbf{RQ3}): \textit{
Synthesized programs can help mitigate the impact of parallelism placement.
}

Consider G1 and G2. As discussed before,
G2's $\allreduce$ (89.7s) is 448.5$\times$ slower than G1 (0.20s).
Synthesized programs have helped bridge the gap:
G2's optimal program is only 330.2$\times$ slower.
However, the performance difference here is significant
and the help is limited.
The case of
H1 and H2 is more interesting:
H1's $\allreduce$ is 1.03$\times$ slower than H2,
but its optimal program is 1.49$\times$ faster than H2!

On the other hand, it is also possible that synthesis
aggravates the impact of parallelism placement.
For example, for I1 and I2,
the performance difference jumps from 
1.10$\times$ for $\allreduce$ to 1.60$\times$ for the optimal program.

\textbf{Result 5} (\textbf{RQ3}): \textit{
For reduction across nodes,
a topology-aware reduction program tends to outperform a single step $\allreduce$,
with speedup on average 1.28$\times$, upto 2.04$\times$.
}

\Cref{table:experiment:2} shows that
when cross-node communication is needed
the optimal program tends to outperform
$\allreduce$.
For example, the speedup is 1.84$\times$ in F2, and 2.04$\times$ in K2. 
For 69$\%$  of all mappings across both systems,
synthesized programs outperform $\allreduce$ by 1.27$\times$ on average.

\begin{figure}[t]
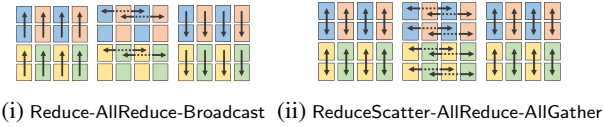

    \centering
    \renewcommand\thesubfigure{\roman{subfigure}}
    \begin{subfigure}[t]{0.21\textwidth}
    \centering
    \includegraphics[width=0.27\linewidth]{images/reduction2.png}
    \includegraphics[width=0.28\linewidth]{images/reduction3.png}
    \includegraphics[width=0.27\linewidth]{images/reduction4.png}
    \caption{\scriptsize{$\reduceno$-$\allreduce$-$\broadcastno$}}
    \label{fig:experiment:reduction:1}
    \end{subfigure}
    \begin{subfigure}[t]{0.255\textwidth}
    \centering
    \includegraphics[width=0.23\linewidth]{images/reduction5.png}
    \includegraphics[width=0.24\linewidth]{images/reduction6.png}
    \includegraphics[width=0.23\linewidth]{images/reduction5.png}
    \caption{\scriptsize{$\reducescatter$-$\allreduce$-$\allgather$}}
    \label{fig:experiment:reduction:2}
    \end{subfigure}
    \caption{Common optimal reduction programs}
    \label{fig:experiment:optimal:reduction}
\end{figure}

We present common optimal reduction programs applied to our running examples
(\Cref{sec:overview}) in \Cref{fig:experiment:optimal:reduction}.
(i)
\Cref{fig:experiment:reduction:1}
first reduces local data to a root device, performs
$\allreduce$ between root devices, and broadcasts the result from the root device
to each device.
(ii) \Cref{fig:experiment:reduction:2}
first performs $\reducescatter$ between local devices, and then
$\allreduce$ between remote devices, and finally $\allgather$ between local devices.
Both reduction programs utilize the topology,
by performing local communication first,
which is often more efficient due to local high bandwidth.
Now, the data to be reduced across nodes in the intermediate step
is significantly smaller.
The final step is again local communication.
Thus, the reduction programs have overall better performance than $\allreduce$.
It turns out that both reduction programs have been recently proposed:
program (i) has been used in \citet{goyal2018accurate, jia2018highly},
and program (ii) has been proposed by \citet{cho2019blueconnect}.

Furthermore, the experiments suggest
that program (ii) is more often the optimal one
and outperforms (i) by a larger speedup.
Specifically, when (i) is optimal, 
it outperforms (ii) by only about $1.1\times$ (up to 1.12$\times$);
when (ii) is optimal, it outperforms (i) by about $1.3\times$ (up to 2.73$\times$).

\begin{figure}[t]
    \centering
    \begin{subfigure}{0.48\textwidth}
    \includegraphics[width=\textwidth,trim=4 4 4 4,clip]{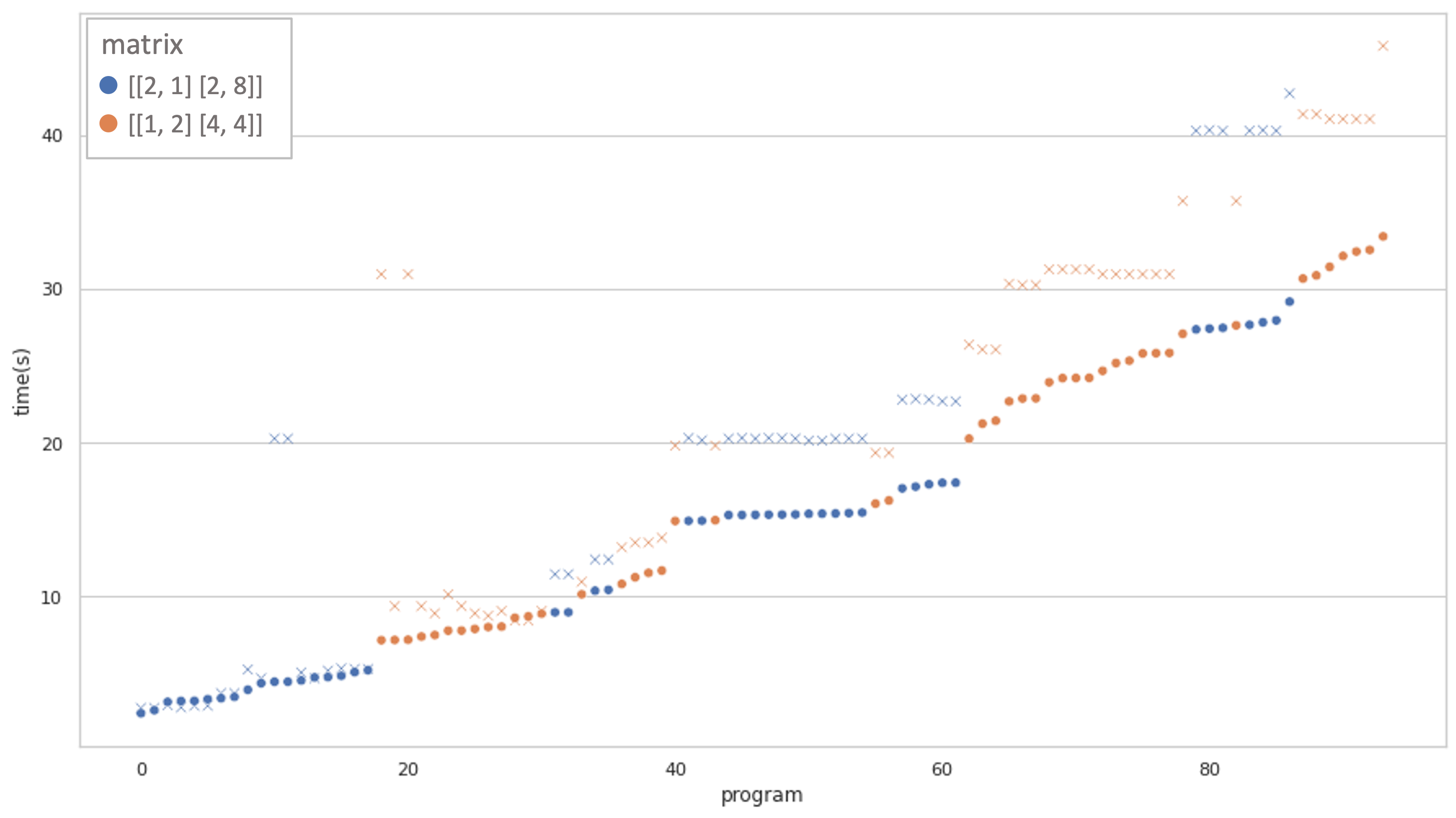}
    \caption{4 nodes of V100 with NCCL Ring and parallelism axes \axes{2&16}, reduction on the 1st axis.
    Synthesis 0.12s, and simulation 0.54s.
    }
    \label{fig:simulation:1}
    \end{subfigure}
    \begin{subfigure}{0.48\textwidth}
    \includegraphics[width=\textwidth,trim=4 4 4 4,clip]{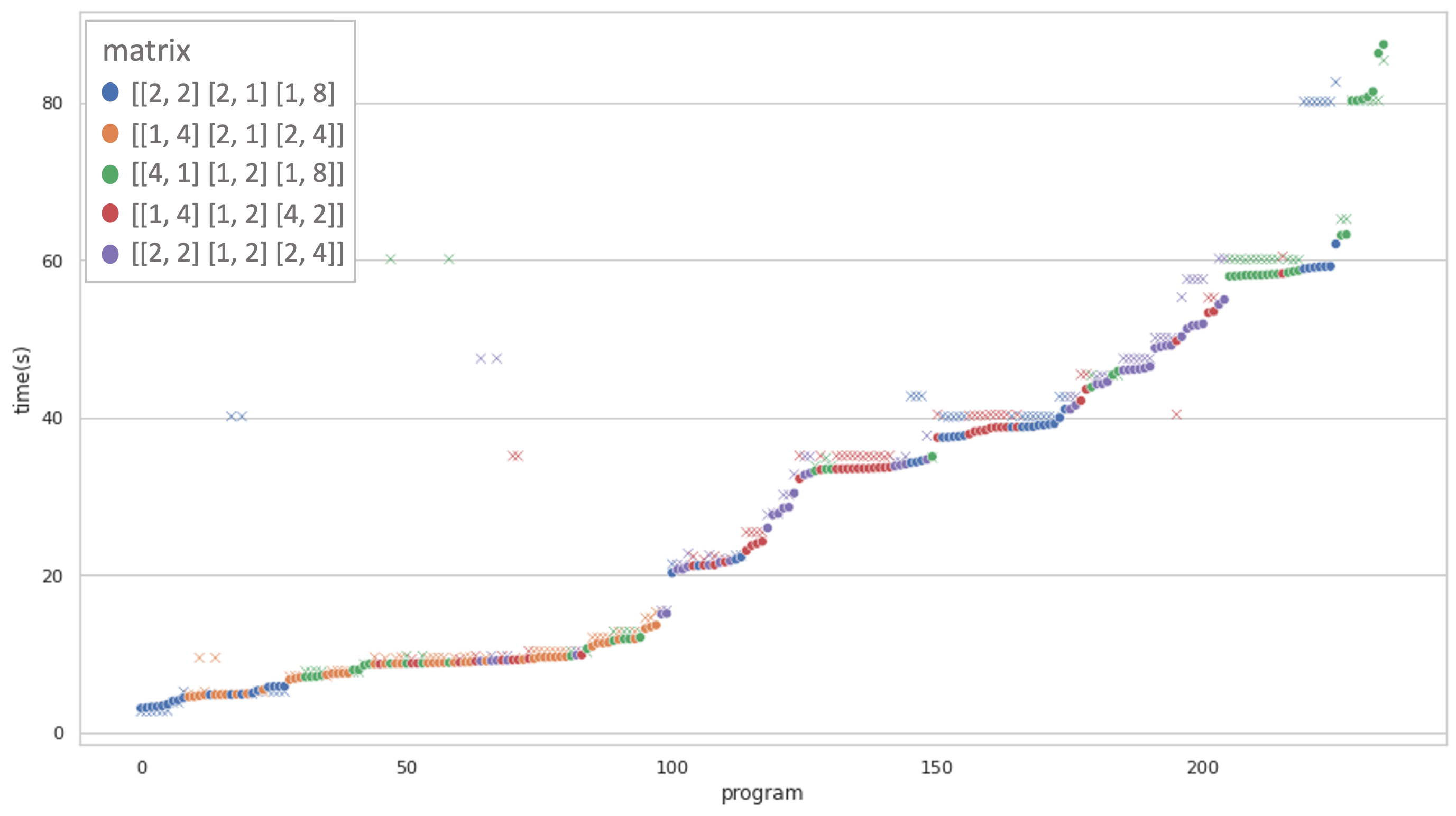}
    \caption{4 nodes of A100 with NCCL Tree and parallelism axes \axes{4&2&8},
    reduction on 0th and 2rd axes. 
    Synthesis 2.86s, simulation 3.09s.
    }
    \label{fig:simulation:4}
    \end{subfigure}
    \caption{Simulation results, in increasing order of experiment time.
    Measurements are '$\bullet$' and solid, and simulations are '$\times$' and translucent.
    Colors denote parallelism matrices.
    }
    \label{fig:simulation}
\end{figure}

\section{Simulation}\label{sec:simulation}

In this section,
we apply simulation to
the system topologies in \Cref{fig:experiment:system},
and
show
that \toolname can identify near-optimal programs, to reduce the need for
evaluation over hundreds or thousands of synthesized mappings and strategies.

\textit{Assumptions.}\,\,We use 100Gbps NICS, which we assume were utilized at
60\%, yielding an effective 8GB/s. For the PCIe switches we assumed 32GB/s. For the V100 NVLink ring we assume 135GB/s in each direction -- an optimistic 90\% of the nominal uni-directional bandwidth (150GB/s)~\cite{v100-dissecting}. For the A100 NVLink switch, we assume 270GB/s uni-directional bandwidth -- again 90\%
of the nominal value (300GB/s in each direction)~\cite{nvidia-nvlink}.

\begin{table}[t]
\small
    \centering
    \setlength\tabcolsep{4pt}
    \begin{tabular}{ccccccc}
        \rowcolor{headrowcolor}
        \hline
              & Top-1  & Top-2  & Top-3 & Top-5  & Top-6 & Top-10  \\ \hline
        A100  & 46.8\% & 71.0\% & 72.6\% & 74.2\% & 90.3\% & 94.4\% \\
        V100  & 60.5\% & 67.1\% & 71.1\% & 76.3\% & 76.3\% & 88.1\% \\
        Total & 52.0\% & 69.5\% & 72.0\%  & 75.0\% & 85.0\% & 92.0\% \\ \hline
    \end{tabular}
    
    \caption{Prediction accuracy.}
    \label{table:predication:accuracy}
\end{table}

\textit{Results.}\,\,\Cref{fig:simulation} shows
that \toolname predictions follow
the same trend as the V100 experiments (\ref{fig:simulation:1}),
and are very close to the absolute performance for A100 (\ref{fig:simulation:4}).
The main reason for reduced absolute accuracy in V100 is the imperfect modeling of cross-domain communication. 
Note that there exist a few programs for which the simulation result is notably slower than the experiments, mostly due to XLA optimizations. For example,
program 10 and 11 (in \ref{fig:simulation:1}) are all 2 steps of $\allreduce$,
which XLA optimizes to a single step $\allreduce$.
We do not extend \toolname with
any optimizations,
as optimized programs are themselves valid {\em synthesizable} programs; in this specific example program 9. \toolname successfully predicts the performance of program 9.

\Cref{table:predication:accuracy} summarizes the accuracy over all
experiments for the two GPU systems. Overall, \toolname delivers 52\% top-1 accuracy, 75\% top-5 accuracy, and 92\% top-10 accuracy.

\section{Related Work}\label{sec:related}

\textit{Parallelism forms.}\,\,
Recent work has explored combinations of parallelism forms.
\citet{jia2018hidden} propose layer-wise parallelism that
allows each network layer to use a different parallelization strategy.
FlexFlow~\cite{jia2019beyond} uses guided randomized search to find a fast
parallelism combination. 
\citet{narayanan2021efficient} combine pipeline, tensor and data parallelism to scale transformer training
to thousands of GPUs, extending previous model parallelism work~\cite{shoeybi2020megatronlm}.
ZeRO-DP~\cite{rajbhandari2020zero}
includes three optimization stages partitioning over optimizer state, gradient, and parameters.
To our best knowledge, no prior work has discussed parallelism placement, and 
typically commit to a specific placement (e.g. model
parallelism within a host, batch parallelism across~\cite{narayanan2021efficient}.)
which can get involved when multiple axes occur. Finally, there exists a rich body of work on operator mapping, e.g. \cite{placeto, spotlight, rl-device-placement}, but the focus there does not include the structured forms of parallelism and reductions we address.

\textit{Program synthesis for communication.}\,\,
For a given topology,
SCCL \cite{cai2021synthesizing} synthesizes optimal collective algorithms as a sequence of sends,
but the work has focused on the single-node setting. 
SCCL takes a more fine-grained system topology as a graph
with bandwidth constraints on GPUs and edges,
and uses an SMT solver to synthesize algorithms.
It is possible for \toolname to use SMT, but we
found that the structure we imposed on the problem already enables fast enumerative
syntax-guided synthesis.
Blink \cite{wang2020blink} performs the synthesis
based on an approximate spanning-tree packing algorithm for intra-node communication,
but always uses program (i) (\Cref{fig:experiment:reduction:2}) for inter-node
communication. 
PLink~\cite{liang2020plink} probes the network locality and groups nodes by
physical affinity, and performs (i) for intra-group reduction.
By contrast, \toolname synthesizes hierarchical strategies
using collectives. 

\textit{Reduction strategies.}\,\,
BlueConnect~\cite{cho2019blueconnect} 
propose program (ii)~(\Cref{fig:experiment:reduction:1})
for hierarchical systems,
which is later generalized by
FlexReduce~\cite{lee2020flexreduce} to asymmetric topologies.
On the other hand, \toolname systematically generates and compares
a wide range of hierarchical reduction strategies for different parallelism placements.

\section{Conclusion}
We have presented a framework to synthesize structured forms of
parallelism mappings and hierarchy-aware reduction strategies. Our tool can be useful for speeding up ML models, and also for establishing projections about communication costs when investigating new system hierarchies.

\bibliography{main}

\newpage
\appendix

\onecolumn

\section{Experiments}

\begin{small}
    \setlength\tabcolsep{4pt}
    \begin{longtable}{M{1.5cm}M{1cm}M{1.2cm}ccccccccccc}
       \rowcolor{headrowcolor}
            Parallelism axes  
            & Reduce axes 
            & Synthesis time 
            &
            \multicolumn{2}{c}{Simulation time}
            &
            \multicolumn{2}{c}{Programs}
            &
            Parallelism matrix 
            &
            \multicolumn{2}{c}{AllReduce}
            &
            \multicolumn{2}{c}{Optimal}
            &
            \multicolumn{2}{c}{Speedup}
            \\
       \rowcolor{headrowcolor}
            & &&
            Ring & Tree
            & Ring & Tree
            &
            & Ring & Tree
            & Ring & Tree
            & Ring & Tree
      \\ \hline
     \multicolumn{6}{l}{2 nodes  each with 16 A100} \\ \hline
     \rowcolor{rowcolor}
     
                   [32]   & [0] & 
                   0.244 
                   & 0.458 & 0.860 &
                   9/47&
                   3/47&
                   [2 16] & 4.74 & 3.57 
                   & 3.18 & 2.83
                   & 1.49 & 1.26
                   \\
                  \rowcolor{rowcolor}
                   
                   [2  16] & [0] 
                   & 0.004
                   & 0.014 & 0.019
                   & 0/6
                   & 1/6
                   & [[1 2] [2 8]] 
                   & 0.12 & 0.17 
                   & 0.12 & 0.15
                   & 1 & 1.13
                   \\
                   
                   
                   & && && &
                   & [[2  1]  [1  16]] 
                   &  36.69 & 37.18
                   & 36.69 & 37.18
                   & 1 & 1
                   \\
                   
                   
                   & [1] 
                   & 0.065
                   & 0.019
                   & 0.208
                   & 10/50
                   & 1/50
                   & [[2  1]  [1  16]] 
                   & 0.18 & 0.22 
                   & 0.18 & 0.20
                   & 1 & 1.1
                   \\
                   
                   
                   & && &&  &
                   & [[1 2] [2 8]] 
                   & 8.44 & 4.82
                   & 4.77 & 4.82
                   & 1.77 & 1
                   \\
                  \rowcolor{rowcolor}

                   [4 8]  & [0] & 0.026 &
                   0.125 & 0.137
                   & 11/50
                   & 5/50
                   & [[1  4]  [2  4]]
                   & 0.15 & 0.20 
                   & 0.15 & 0.17 
                   & 1 & 1.18
                   \\
                   
                   &&&&&&&
                   [[2  2]  [1  8]]
                   & 28.91 & 19.61
                   & 18.48 & 18.5
                   & 1.56 & 1.06
                   \\
                   
                   & [1] &  0.030
                   & 0.135 & 0.149
                   & 12/50
                   & 3/50
                   & [[2  2]  [1  8]] 
                   & 0.17  & 0.21 
                   & 0.17 & 0.18
                   & 1 & 1.17
                   \\
                   
                   & && &  &&
                   & [[1  4]  [2  4]] 
                   & 16.04 & 9.15 
                   & 8.97 & 9.01
                   & 1.79 & 1.02
                   \\
                   
                  \rowcolor{rowcolor}
                   
                   [8 4]  & [0] & 0.033
                   & 0.136 & 0.151
                   & 14/50
                   & 2/50
                   & [[1  8]  [2  2]] & 0.17 & 0.21 
                   & 0.17 & 0.19
                   & 1 &  1.11
                   \\
                   &&&&&& &
                    [[2  4]  [1  4]]  
                    & 16.84 & 9.26 
                    & 9.19 & 9.10
                    & 1.83 & 1.02
                    \\
                    
                    & [1] & 0.026 
                    & 0.124 & 0.136
                   & 11/50
                   & 7/50
                   & [[2  4]  [1  4]] 
                   & 0.15 &  0.20 
                   & 0.15 & 0.17
                   & 1 & 1.18
                   \\
                   
                   &&&&&&&
                   [[1  8]  [2  2]] 
                   & 28.78 & 18.93 
                   & 18.48 & 18.00
                   & 1.56 & 1.05
                   \\
                   
                  \rowcolor{rowcolor}
                   
                   [16 2] & [0] & 0.067 
                   & 0.195
                   & 0.214
                   & 10/50
                   & 2/50
                   & [[1  16]  [2  1]] & 0.18 & 0.22 
                   & 0.18 & 0.19
                   & 1 & 1.16
                   \\
                   & & & & & & &
                   [[2  8]  [1  2]] & 8.86 & 5.34 
                   & 5.05 & 5.21
                   & 1.75 & 1.02
                   \\
                    
                   & [1]  & 0.005 
                   & 0.011 
                   & 0.012
                   & 0/6
                   & 1/6
                   & [[2  8]  [1  2]] & 0.11 & 0.17 
                   & 0.11 & 0.14
                   & 1 & 1.21
                   \\
                   & & & & & & &
                   [[1  16]  [2  1]]
                   & 36.84 & 36.82 
                   & 36.84 & 36.82 
                   & 1 & 1
                   \\
                   
      \\ \hline
     \multicolumn{6}{l}{4 nodes  each with 16 A100} \\ \hline
                  \rowcolor{rowcolor}
                   
                   [64]  & [0] & 1.161
                   & 1.868 & 2.01
                   & 6/47
                   & 5/47
                   & [[4 16]]
                   & 5.18 & 5.75 
                   & 4.29 & 4.74
                   & 1.21 & 1.21
                   \\
                  \rowcolor{rowcolor}
                   
                   [2 32] & [0] & 0.006 & 
                   0.019 & 0.022
                   & 0/6
                   & 1/6
                   & [[1  2]  [4  8]]
                   & 0.12& 0.17 
                   & 0.12& 0.15
                   & 1 & 1.13
                   \\
                   
                   && & && & &
                   [[2  1]  [2  16]] 
                   & 37.16 & 36.94 
                   & 37.04 & 36.94 
                   & 1.003 & 1 
                   \\
                   
                   & [1] & 0.463
                   & 0.160 & 1.27
                   & 18/94
                   & 10/94
                   & [[2  1]  [2  16]]
                   & 4.81 & 3.41 
                   & 3.05 & 3.07
                   & 1.58 & 1.11
                   \\
                   
                   & && & &&&
                   [[1  2]  [4  8]] 
                   & 8.74 & 9.89 
                   & 6.91 & 8.00
                   & 1.26 & 1.24
                   \\
                  \rowcolor{rowcolor}
                   
                   [4 16] & [0] & 0.043 &
                   0.259 & 0.287 &
                   12/53
                   & 10/53
                   & [[1  4]  [4  4]] 
                   & 0.15 & 0.20
                   & 0.15 & 0.17
                   & 1 & 1.18
                   \\
                   & && & && &
                   [[2  2]  [2  8]] 
                   & 28.77 & 19.81 
                   & 18.24 & 18.55
                   & 1.57 & 1.07
                   \\
                   & & & & & & &
                   [[4  1]  [1  16]] 
                   & 56.13 & 89.70
                   & 55.99 & 56.13
                   & 1.003 & 1.60
                   \\
                   
                   &
                   [1] & 0.133 &
                   0.619 & 0.704&
                   21/97
                   & 9/97
                   & [[4  1]  [1  16]] 
                   & 0.18 & 0.22 
                   & 0.18 & 0.19 
                   & 1 & 1.16
                   \\
                   & && &  &&&
                   [[2  2]  [2  8]] 
                   & 8.39 & 4.99 
                   & 4.81 & 4.82 
                   & 1.74 & 1.04
                   \\
                   & & && &&&
                   [[1  4]  [4  4]] 
                   & 17.70 & 19.03
                   & 13.38 & 14.85
                   & 1.32 & 1.28
                   \\
                  \rowcolor{rowcolor}
                   
                   [8 8]  & [0] & 0.091 & 
                   0.512 & 0.590&
                   20/97
                   & 7/97
                   & [[1  8]  [4  2]] 
                   & 0.17 & 0.21 
                   & 0.17 & 0.18 
                   & 1 & 1.17
                   \\
                    & &&& & &&
                    [[2  4]  [2  4]] 
                    & 16.52 & 9.18
                    & 9.21 & 9.18
                    & 1.79 & 1
                    \\
                    & &&& & &&
                    [[4  2]  [1  8]] 
                    & 34.05 & 41.23 
                    & 28.86 & 29.79
                    & 1.18 & 1.38
                    \\
                    & [1] & 0.084 & 
                    0.508 & 0.598 
                    & 19/97
                    & 17/97
                    & [[4  2]  [1  8]] 
                    & 0.17 & 0.21 
                    & 0.17 & 0.18
                    & 1 & 1.17
                     \\
                    & & &&& &&
                    [[2  4]  [2  4]] 
                    & 15.68 & 9.43 
                    & 8.92 & 9.22
                    & 1.76 & 1.02
                    \\
                    
                    & & & &&& &
                    [[1  8]  [4  2]]
                    & 33.92 & 41.06 
                    & 27.93 & 29.36
                    & 1.21 & 1.40
                    \\
                  \rowcolor{rowcolor}
                    
                   [16 4] & [0] & 0.149 &
                   0.631 & 0.712 &
                    21/97
                    & 17/97
                    & [[1  16]  [4  1]] 
                    & 0.18 & 0.22 
                    & 0.18 & 0.2
                    & 1 & 1.1
                    \\
                    & && && &&
                    [[2  8]  [2  2]] 
                    & 8.81 & 5.42
                    & 5.01 & 5.25
                    & 1.76 & 1.03
                    \\
                    & && &&  &&
                    [[4  4]  [1  4]] 
                    & 18.30 & 20.13 
                    & 14.13 & 14.90
                    & 1.30 & 1.35
                    \\
                    & [1] & 0.042 & 
                    0.261 & 0.297
                    & 13/53
                    & 5/53
                    &[[4  4]  [1  4]] 
                    & 0.15 & 0.20 
                    & 0.15 & 0.18
                    & 1 &  1.11
                    \\
                    & && & &&&
                    [[2  8]  [2  2]] 
                    & 28.68 & 18.47 
                    & 19.09 & 18.41 
                    & 1.50 & 1.003
                    \\
                     & && & &&&
                    [[1  16]  [4  1]] 
                    & 57.13 & 85.22
                    & 56.23 & 55.63
                    & 1.02 & 1.53
                    \\
                  \rowcolor{rowcolor}
                    
                   [32 2] & [0] & 0.483 &
                   1.183 & 1.30&
                   20/94
                   &
                   15/94
                   & [[2  16]  [2  1]] 
                   & 4.74 & 3.99 
                   & 3.14 & 3.13
                   & 1.51 & 1.27
                   \\
                   & && & &&&
                   [[4  8]  [1  2]] 
                   & 9.37 & 10.41 
                   & 7.23 & 7.71
                   & 1.30 & 1.35
                   \\
                   & [1] & 0.008 &  
                   0.020 &0.022&
                    0/6
                  & 1/6
                   & [[4  8]  [1  2]] 
                   & 0.11 & 0.17 
                   & 0.11 & 0.15
                   & 1 & 1.13
                   \\
                    & & && && &
                   [[2  16]  [2  1]] 
                   & 37.18 & 37.10 
                   & 37.18 & 37.10 
                   & 1 & 1
                   \\
                  \rowcolor{rowcolor}
                   
                   [16 2 2] & [0 2] & 0.968 & 
                   2.36 & 2.55&
                   25/188
                   & 21/188
                   & [[1  16]  [2  1]  [2  1]] 
                   & 4.79 & 3.69 
                   & 4.63 & 2.71
                   & 1.03 & 1.36
                   \\
                 & && & &&&
                 [[2  8]  [2  1]  [1  2]] 
                 & 4.91 & 3.97 
                 & 3.10 & 2.93 
                 & 1.58 & 1.35
                 \\
                 & & && &&&
                 [[2  8]  [1  2]  [2  1]] 
                 & 9.05 & 10.29 
                 & 9.03 & 9.46
                 & 1.002 & 1.09
                 \\
                 & & && &&&
                 [[4  4]  [1  2]  [1  2]] 
                 & 9.14 & 10.32 
                 & 7.08 & 7.87
                 & 1.29 & 1.31
                 \\
                  \rowcolor{rowcolor}
                  
                     [8 2 4] & [0 2] & 1.107 &
                     2.88 & 3.08&
                     28/235&
                     22/235
                     & [[1  8]  [2  1]  [2  2]] 
                     & 4.80 & 3.62 
                     & 4.64 & 2.67
                     & 1.03 & 1.36
                     \\
                     & & &&&&
                     & [[2  4]  [2  1]  [1  4]] 
                     & 4.82 & 3.87 
                     & 3.12 & 3.08
                     & 1.54 & 1.26
                     \\
                     & & &&&&
                     & [[2  4]  [1  2]  [2  2]] 
                     & 8.91 & 9.68 
                     & 8.91 & 9.29 
                     & 1 & 1.04
                     \\
                     & & &&&&
                     & [[4  2]  [1  2]  [1  4]] 
                     & 9.19 & 10.24 
                     & 7.02 & 7.69 
                     & 1.31 & 1.33
                     \\
                     & & &&&&
                     & [[1  8]  [1  2]  [4  1]] 
                     & 9.21 & 10.37 
                     & 5.50 & 8.72
                     & 1.7 & 1.19
                     \\
                     \rowcolor{rowcolor}
                  
                     [4 2 8] & [0 2] & 1.143 &
                     2.86 & 3.09&
                     32/235&
                     24/235&
                     [[2  2]  [2  1]  [1  8]] 
                     & 4.74 & 3.54
                     & 3.04 & 2.99
                     & 1.56 & 1.18
                     \\
                     & & & &&&&
                     [[1  4]  [2  1]  [2  4]] 
                     & 4.77 & 3.77
                     & 4.48 & 3.54
                     & 1.06 & 1.06
                     \\
                     & & & &&&&
                     [[4  1]  [1  2]  [1  8]] 
                     & 8.73 & 9.81
                     & 7.00 & 8.12
                     & 1.25 & 1.21
                     \\
                     & & & &&&&
                     [[1  4]  [1  2]  [4  2]] 
                     & 9.12 & 9.98
                     & 8.65 & 8.80
                     & 1.05 & 1.13
                     \\
                     & & & &&&&
                     [[2  2]  [1  2]  [2  4]] 
                     & 9.12 & 10.36
                     & 9.02 & 9.77
                     & 1.01 & 1.06
                     \\
                     \rowcolor{rowcolor}
                    
                     [2 2 16] & [0 2] & 0.927 &
                     2.32 & 2.51&
                     29/188&
                     16/188
                     & [[2  1]  [2  1]  [1  16]] 
                     & 4.82 & 3.91 
                     & 2.99 & 3.00
                     & 1.61 & 1.30
                     \\
                     & & & &&&&
                     [[1  2]  [2  1]  [2  8]] 
                     & 5.28 & 4.29 
                     & 4.77 & 3.66
                     & 1.11 & 1.17
                     \\
                     & & & &&&&
                     [[2  1]  [1  2]  [2  8]] 
                     & 9.32 & 9.61
                     & 7.10 & 7.97
                     & 1.31 & 1.21
                     \\
                     & & & &&&&
                     [[1  2]  [1  2]  [4  4]] 
                     & 9.81 & 9.79
                     & 9.81 & 9.37
                     & 1 & 1.04
                     \\
                     \\
                   \hline
     \multicolumn{6}{l}{2 nodes  each with 8 V100} \\ \hline
                    \rowcolor{rowcolor}

                   [16]    & [0] & 0.058 & 0.158& 0.178
                   & 12/47
                   & 0/47
                   & [[2  8] ] & 4.58 & 2.30 
                   & 2.41 & 2.3 & 1.90 & 1
                   \\
                    \rowcolor{rowcolor}
                   
                   [2 8]   & [0] & 0.0035&0.014& 0.134&
                   0/6&
                   0/6&
                   [[1  2]  [2  4]] &9.37 & 8.44 
                  & 9.37 & 8.44
                   & 1 & 1
                   \\
                   &&& & & &&
                   [[2  1]  [1  8]] 
                   &14.53 & 14.55
                   &14.53 & 14.55
                   & 1 & 1
                   \\
                   
                   & [1] & 0.0257 & 0.206& 0.128&
                   6/50&
                   2/50&
                     [[2  1]  [1  8]] 
                     & 0.28 & 0.40 
                     & 0.28 & 0.30 
                     & 1 & 1.33
                     \\
                    && && &&&
                    [[1  2]  [2  4]] 
                    & 6.59 & 3.70 
                    & 6 & 3.70 
                    & 1.10 & 1
                    \\
                    \rowcolor{rowcolor}
                    
                    [4 4]   & [0] & 0.0181 & 0.243& 0.161&
                   0/50&
                   8/50&
                    [[1  4]  [2  2]] 
                    & 12.55 & 13.01 
                    & 12.55 & 13.01 
                    & 1 & 1
                    \\
                    && && &&&
                    [[2  2]  [1  4]] 
                    & 13.11 & 16.11 
                    & 13.11 & 13.5
                    & 1 & 1.19
                    \\
                    & [1] & 0.0182 & 0.181& 0.116&
                    10/50&
                    2/50&
                    [[2  2]  [1  4]] 
                    & 2.96 & 0.43 
                    & 2.29 & 0.43
                    & 1.29 & 1
                    \\
                    &&& & &&&
                    [[1  4]  [2  2]] 
                    & 10.95 & 7.42 
                    & 7.52  & 7.26
                    & 1.46 & 1.02
                    \\
                    \rowcolor{rowcolor}
                    
                    [8 2]   & [0] & 0.0272 & 0.137& 0.304&
                   1/50&
                   4/50&
                    [[1  8]  [2  1]] 
                    & 0.28 & 0.40 
                    & 0.28 & 0.30 
                    & 1 & 1.33
                    \\
                   & &&& &&&
                   [[2  4]  [1  2]] 
                   & 14.24 & 15.47  
                   & 14.2 & 14.48 & 1.003 & 1.07
                   \\
                    & [1] & 0.0036 & 0.009& 0.023 &
                   0/6&
                   1/6&
                    [[2  4]  [1  2]] 
                    & 0.32 & 0.33 
                    & 0.32 & 0.33 
                    & 1 &1
                    \\
                   & && &&&&
                   [[1  8]  [2  1]] 
                   & 14.51 & 14.81 
                   & 14.51& 14.47
                   & 1 & 1.02
                   \\
                   \\
                   \hline
     \multicolumn{6}{l}{4 nodes  each with 8 V100} \\ \hline
                    \rowcolor{rowcolor}
                    
                  [32]    & [0] & 0.209 &  0.507& 0.770&
                  11/47 &
                  7/47 &
                  [[4  8] ] 
                  & 4.83 & 4.57 
                  & 4.83 & 3.66
                  & 1 & 1.25
                  \\
                    \rowcolor{rowcolor}
                    
                  [2 16]   & [0] & 0.0043 & 0.024& 0.027&
                  0/6 &
                  1/6 &
                  [[1  2]  [4  4]] 
                  & 9.17 & 8.42 
                  & 9.17 & 8.42 
                  & 1 & 1
                  \\
                   & && &&&&
                  [[2  1]  [2  8]] 
                  & 14.47 & 14.52
                  & 14.47 & 14.51
                  & 1 & 1.0007
                  \\
                  
                   [2 16]   & [1] & 0.121 & 0.536& 0.621&
                  10/94 &
                  7/94 &
                   [[2  1]  [2  8]] 
                   & 4.37 & 2.26 
                   & 2.42 & 2.25
                   & 1.81 & 1.004
                   \\
                   & & &&&&&
                   [[1  2]  [4  4]] 
                   &  7.18 & 8.10
                   &  7.16 & 7.10
                   &  1.003 & 1.14
                   \\
                   
                   [4 8]   & [0] & 0.027 & 0.303& 0.356&
                  1/53 &
                  9/53 &
                  [[1  4]  [4  2]]  
                  & 12.60 & 13.00 
                  & 12.60 & 13.00 
                  & 1 & 1
                  \\
                   & && &&&&
                  [[2  2]  [2  4]] 
                  & 13.69 & 16.07
                  & 13.69 & 13.56
                  & 1 & 1.19
                  \\
                   & && &&&&
                  [[4  1]  [1  8]] 
                  & 22.21 & 30.55 
                  & 22.04 & 21.49 & 1.001 & 1.42
                  \\
                    \rowcolor{rowcolor}
                    
                  [4 8]   & [1] & 0.096 & 0.421& 0.498&
                  15/97 &
                  11/97 &
                   [[4  1]  [1  8]]   
                   & 0.28 & 0.40 
                   & 0.28 & 0.30 
                   & 1 & 1.33
                   \\
                   & & &&&&&
                   [[2  2]  [2  4]]  
                   & 6.60 & 3.82 
                   & 5.78 & 3.74
                   & 1.14 & 1.02
                   \\
                   & & && &&&
                   [[1  4]  [4  2]]
                   & 12.94 & 15.47 
                   & 11.21 & 12.12
                   & 1.15 & 1.28
                   \\
                    \rowcolor{rowcolor}
                    
                  [8 4]   & [0] & 0.103 & 0.579& 0.701&
                  2/97 &
                  13/97 &
                  [[1  8]  [4  1]] 
                  & 0.28  & 0.39
                  & 0.28 & 0.3
                  & 1 & 1.3
                  \\
                   & & &&&&&
                  [[2  4]  [2  2]] 
                  & 14.25 & 15.48 
                  & 14.23 & 14.49
                  & 1.001 & 1.07
                  \\
                   & & &&&&&
                  [[4  2]  [1  4]] 
                  & 14.84 & 19.90 
                  & 1.84 & 15.33
                  & 1 & 1.30
                  \\
                     & [1] & 0.0435 & 0.213& 0.241&
                  11/53 &
                  2/53 &
                  [[4  2]  [1  4]] 
                  & 2.96 & 0.43 
                  & 2.29 & 0.43 
                  & 1.29 & 1
                  \\
                   & & &&&&&
                  [[2  4]  [2  2]] 
                  & 10.98 & 7.34 
                  & 7.58 & 7.34 
                  & 1.45 & 1
                  \\
                   & & &&&&&
                  [[1  8]  [4  1]] 
                  & 21.74 & 30.42 
                  & 21.74 & 21.71
                  & 1 & 1.40
                  \\
                  
                    \rowcolor{rowcolor}
                  
                   [16 2]   & [0] & 0.168 & 0.597& 0.719
                   & 12/94
                   & 7/94
                   & [[2  8]  [2  1]] 
                   & 4.47 & 2.25 
                   & 2.44 & 2.25
                   & 1.83 & 1
                   \\
                   & & &&&&&
                   [[4  4]  [1  2]] 
                   & 15.00 & 17.58
                   & 14.99 & 15.01
                   & 1.0007 & 1.17
                   \\
                    
                    & [1] & 0.004 & 0.016 & 0.019&
                   0/6 &
                   1/6 &
                   [[4  4]  [1  2]] 
                   &  0.32 & 0.33 
                   &  0.32 & 0.33 
                   &  1 & 1
                   \\
                   & & &&&&&
                   [[2  8]  [2  1]] 
                   & 14.53 & 14.89
                   & 14.53 & 14.66
                   & 1 & 1.02
                   \\
                    \rowcolor{rowcolor}
                    
                   [2 2 8]   & [0 2] & 0.229  & 1.105&&
                   14/188 &&
                   [[1  2]  [2  1]  [2  4]] 
                   & 4.29 
                   &
                   & 4.29 
                   &
                   & 1
                   \\
                   & & &&&&&
                   [[2  1]  [2  1]  [1  8]]  
                   & 4.57 
                   &
                   & 2.4
                   &
                   & 1.90
                   \\
                   & & &&&&&
                   [[2  1]  [1  2]  [2  4]] 
                   & 7.17 
                   &
                   & 6.95
                   &
                   & 1.03
                   \\
                   & & &&&&&
                   [[1  2]  [1  2]  [4  2]] 
                   & 9.41 
                   &
                   & 9.41 
                   &
                   & 1
                   \\
                    \rowcolor{rowcolor}
                    
                     [8 2 2]   & [0 2] & 0.243 & 1.184&&
                    17/188 &&
                    [[1  8]  [2  1]  [2  1]] 
                    &  4.40 
                    &
                    &  4.40 
                    &
                    &  1
                    \\
                    & & &&&&&
                    [[2  4]  [2  1]  [1  2]] 
                    &  4.80 
                    &
                    &  2.35
                    &
                    &  2.04
                    \\
                    & & &&&&&
                    [[4  2]  [1  2]  [1  2]] 
                    &  9.36 
                    &  
                    &  9.35
                    &  
                    &  1.001
                    \\
                    & & &&&&&
                    [[2  4]  [1  2]  [2  1]] 
                    &  15.02
                    &
                    &  14.95
                    &
                    &  1.005
                    \\
                   \bottomrule 
    \end{longtable}
\end{small}

\twocolumn

\section{Proof of expressiveness between synthesis hierarchies}

Recall our definitions of synthesis hierarchies: 

(a) System hierarchy ({\small{$\begin{bmatrix}1&2&2&4\end{bmatrix}$}}) \\
(b) Column-based parallelism factors ({\small{$\begin{bmatrix} 1 & \makecmcolor{1} & 1 & \makecmcolor{2} & 2 & \makecmcolor{1} & 2 & \makecmcolor{2}\end{bmatrix}$}}) \\
(c) Row-based parallelism factors ({\small{$\begin{bmatrix}1 & 1 & 2 & 2 & \makecmcolor{1} & \makecmcolor{2} & \makecmcolor{1} & \makecmcolor{2} \end{bmatrix}$}})\\
(d) Reduction axis parallelism factors ({\small{$\begin{bmatrix}\makecmcolor{1} & \makecmcolor{2} & \makecmcolor{1} & \makecmcolor{2} \end{bmatrix}$}})

Note that while we can assume all system hierarchies (i.e., (a), and thus (b) and (c)) start with a level of 1, e.g., [(rack, 1), (server, 2), (CPU, 2), (GPU, 4)], it may not be the case for (d). To make it a complete synthesis hierarchy,
we will append a (root, 1) as the root of (d).

Our goal is to prove

\thmexpressive*

We split our goal into three parts:
(1) (b) $\geq$ (a) (\Cref{sec:appendix:part1} \Cref{lemma:part1});
(2) (c) $\geq$ (b) (\Cref{sec:appendix:part2} \Cref{lemma:part2});
(2) (d) $\geq$ (c) (\Cref{sec:appendix:part3} \Cref{lemma:part3}),
and prove them separately.

During the proof, 
we often use parallelism factors ($x$)
and their corresponding levels ($e$) interchangeably.

\subsection{Part 1}
\label{sec:appendix:part1}

\begin{lemma}
\label{lemma:part1}
(b) $\geq$ (a).
\end{lemma}

\proof

This lemma is intuitive, as (b) expands (a), so can be used
to express any communication patterns that can be formed by (a).

Suppose (a) is \axes{x_0 & x_1 & \cdots & x_n},\\
and the parallelism axes has size $0..m$,\\
then (b) is \axes{x_{00} & \cdots & x_{0m} & x_{10} & \cdots & x_{1m} & \cdots & x_{n0} & \cdots & x_{nm} },\\
with
$\mathlarger{\mathlarger{\prod}}_{j=0}^{m}x_{ij} = x_i$.

Now we show that every reduction instruction given by (a),
a reduction instruction can be formed by (b) expressing the same reduction.

\begin{description}
    \item[Case 1]: (a) uses ($x_i$, $\insidegroup$, $\op$).
    
    Then (b) can use ($x_{i0}$, $\insidegroup$, $\op$)
    to express the same reduction.
    
    To see why this is true, note that $x_{ij}$, forall $j$,
    represents the same system hierarchy as $x_i$.
    So $x_{i0}$ and $\insidegroup$ form the same device
    groups as $x_i$ and $\insidegroup$.
    
    \item[Case 2]: (a) uses  ($x_i$, $\parallel{x_j}$, $\op$).
    
    Then (b) can use  ($x_{i0}$, $\parallel{x_{jm}}$, $\op$)
    to express the same reduction. 
    
    Here, the use of $x_{i0}$ ensures that we form the same reduction groups, and then we use
    $x_{jm}$ (instead of $x_{j0}$) to get the first (second/etc) device
    from each reduction group that connects to the same $x_{jm}$.
    As $x_{jm}$ is the last parallelism factor for $x_j$,
    this makes sure that we only connect devices that
    connect to the same $x_j$.
    
    \item[Case 3]: (a) uses ($x_i$, $\master{x_j}$, $\op$).
    
    Then (b) can use  ($x_{i0}$, $\master{x_{jm}}$, $\op$)
    to express the same reduction. The case is similar as the above one.
    
\end{description}
\qed

\subsection{Part 2}
\label{sec:appendix:part2}

\subsubsection{Semantically valid reduction}

\begin{lemma}[Parallel reduction]
\label{lemma:parallel:observation}
Consider a synthesis hierarchy 
\axes{x_0 & x_1 & \cdots & x_j & \cdots & x_i & \cdots x_{n} },
and the device groups formed by
($x_i$, $\parallel{x_j}$, $\op$), we have:
\begin{enumerate}
\item 
a device group to be reduced has size $x_{j+1} \times ... \times x_i$,
and we have
$x_0 \times x_1 \times ... \times x_{j} \times x_{i+1} \times ... \times x_n $ number of such groups.

\item
a device group has been partitioned over $x_{j+1}$, ..., $x_{i}$.
\end{enumerate}
\end{lemma}

\proof

We can derive the observation from the semantics of $\parallelno$.
In particular,
we first form reduction groups for devices connected to the same $x_i$.
So each reduction group has size $size_1 = x_{i+1} \times ... \times x_n$.

Then, because of $\parallelno$,
we reduce all first (second/etc) devices in the reduction group if they are
connected to the same $x_j$.
Since each $x_j$ owns $size_2 = x_{j+1} \times ... \times x_i$ different
reduction groups, the device group we form is
exactly of $size_2$.
And for each $x_j$, we have $size_1$
of such device groups.
Since we have $size_3 = x_0 \times ... \times x_{j}$ different $x_j$,
we have in total $size_3 \times size_1$ device groups.

The second observation is similar. In particular, since device groups
are formed by each $x_j$, which owns $size_2$ different $x_i$s with
different $x_{j+1}, ..., x_{i}$, the device groups we form are exactly partitioned
over $x_{j+1}, ..., x_{i}$.

\qed

\begin{lemma}[Partitioning over reduction axes]
\label{lemma:partition:over:reduction:axes}
Given reduction axes,
for a reduction instruction to be semantically valid,
all device groups to be reduced must
only be partitioned over the reduction axes.
\end{lemma}

\proof

Suppose we want to reduce over reduction axes A, and we reduce between $d_j$
and $d_j$ that have been, possibly among others, partitioned over B (different from A).
Then, after reduction both devices contain data that differs in B.
Now the desired final state (where data should only be reduced if they have different A)
becomes unreachable for both devices,
as we can never recover the state of the device
since according to the semantics,
once a row has grown it will never get reduced back.
\qed

\begin{corollary}[Semantically valid parallel reduction]
\label{lemma:valid:parallel:reduction}
Consider a synthesis hierarchy 
\axes{x_0 & x_1 & \cdots & x_j & \cdots & x_i & \cdots x_{n} },
given some reduction axes, a reduction instruction
$(x_i, \parallel{x_j}, \op)$
is only semantically valid
if all of $x_{j+1}, ..., x_{i}$
are either 1, or on the reduction axes.
\end{corollary}

\proof

By \Cref{lemma:parallel:observation}, we know each device group
is partitioned over $x_{j+1}, ..., x_{i}$.
However,
\Cref{lemma:partition:over:reduction:axes} shows that
for each device group to be semantically valid,
they can only be partitioned over the reduction axes.
Therefore, all $x$s in $x_{j+1}, ..., x_{i}$.
should either be 1, or on the reduction axes.
\qed

\begin{lemma}[Semantically valid InsideGroup reduction]
\label{lemma:valid:insidegroup}
Consider a synthesis hierarchy 
\axes{x_0 & x_1 & \cdots & x_j & \cdots & x_i & \cdots x_{n} },
given some reduction axes, a reduction instruction
$(x_i, \insidegroup, \op)$
is only semantically valid
if all of $x_{i+1}, ..., x_{n}$
are either 1, or on the reduction axes.
\end{lemma}

\proof

Similar as \Cref{lemma:parallel:observation},
except in this case we know that we are reducing devices over 
$x_{i+1}, ..., x_{n}$.

\qed

\begin{lemma}[Semantically valid master reduction]
\label{lemma:valid:master}
Consider a synthesis hierarchy 
\axes{x_0 & x_1 & \cdots & x_j & \cdots & x_i & \cdots x_{n} },
given some reduction axes, a reduction instruction
$(x_i, \master{x_j}, \op)$
is only semantically valid
if all of $x_{j+1}, ..., x_{n}$
are either 1, or on the reduction axes.
\end{lemma}

\proof Note that in this case we require all parallelism factors up until $x_n$ (as with $\insidegroup$, rather than $x_i$ as with $\parallelno$).
The trickiness here is that the case of $\masterno$ is different from $\parallelno$: while in $\parallelno$ we know that we will reduce in parallel everything that is not in the range of $\parallelno$ (\Cref{lemma:parallel:observation}), with $\masterno$ we reduce only the \textit{first} reduction group. So it is important to guarantee that we form exactly the same first inner reduction groups.

We prove the result by contradiction.
Suppose $x_{j+1}, ..., x_{n}$ contains a parallelism factor $x'$ that is not 1 nor is on the reduction axes.
Then there are two possibilities.

(1) $x'$ is part of $x_{j+1}, ..., x_{i}$, i.e., it is covered by the reduction. Then we will reduce devices of different $x'$. According to \cref{lemma:partition:over:reduction:axes}, the reduction is invalid.

(2) $x'$ is part of $x_{i+1}, ..., x_{n}$, i.e., it is not covered
by the reduction, but it affects the reduction groups we form.
Suppose reduction axes are A, and $x'$ is on level B, with $B \neq A$.

Then since $x'$ is part of $x_{i+1}, ..., x_{n}$,
we can find within $x_{i+1}, ..., x_{n}$ a device $d_{0B}$,
that has only different B with the very first device $d_0$.

Suppose $d_0$ reduces with some device $d_{1}$ in this master reduction.
Since the reduction is valid, $d_0$ and $d_1$ differ only in A.

Then in the same way we find $d_{dB}$ to $d_{0}$,
we can find a device $d_{1B}$, that is only different with $d_1$ in B.

However, note that this is a master reduction, and $d_0$ and $d_{0B}$
belong to the same $x_{i+1}, ..., x_{n}$ group,
so the master reduction will only reduce $d_0$ and $d_1$,
but only $d_{0B}$ and $d_{1B}$.

Now we can show that the final desired state becomes unreachable.
In particular,  note that $d_{0B}$ and $d_{1B}$ will never get reduced:
every reduction that reduces $d_{0B}$ and $d_{1B}$ will reduce $d_0$
and $d_1$ as well (\Cref{fig:composition} is useful here). But since $d_0$ and $d_1$ has been reduced already,
re-reducing the devices is an invalid reduction step.
So we will never
be able to reduce $d_{0B}$ and $d_{1B}$ again. And thus
the master reduction is invalid.

\qed

\begin{lemma}
\label{lemma:part2}
(c) $\geq$ (b).
\end{lemma}

\proof

\begin{description}
    \item[Case 1] (b) has  $(e_2, \parallel{e_1}, \op)$.
    
    Based on \Cref{lemma:valid:parallel:reduction},
    all non-reduction parallelism factors column-wisely
    between $e_1$ (exclusive) and $e_2$ (inclusive) can only be $1$.
    
    Now we construct a reduction instruction for (c) that expresses the same reduction.
    Suppose the reduction step in (b) covers parallelism factors $e_i,...,e_j$ on the reduction axis.
    Note that if the reduction step in (b) does not cover any parallelism factors, that means it forms groups of one device, and in that case
    this does not form a reduction and won't be generated by \toolname. 
    
    Let $e_1'$ be the level right
    before $e_i$ row-wisely in the synthesis hierarchy (c),
    and let $e_2'$ be $e_j$. Then $(e_2', \parallel{e_1'}, \op)$ is a desired reduction instruction. Indeed, we can derive from \Cref{lemma:parallel:observation} that
    this reduction instruction forms the same device groups as (b).
    
    The example from the paper is repeated  below, with the reduction axis highlighted.
    
\noindent
\begin{minipage}{0.03\textwidth}
\quad
\end{minipage}
\begin{minipage}{0.18\textwidth}
\begin{small}
\begin{gather*}
\begin{bmatrix} 
    x_{0,0}  & 1 & 1 & x_{0,3} \\
    \mypoint{e1}{}{\bm{{x_{1,0}}}} & 1 & 1 & x_{1,3} \\
    \makecmcolor{x_{2,0}} & \makecmcolor{x_{2,1}} & \makecmcolor{x_{2,2}} & \makecmcolor{x_{2,3}} \\
    1 & 1 & \mypoint{e2}{}{\bm{1}} & x_{3,3} \\
\end{bmatrix}
\end{gather*}
 \begin{tikzpicture}[remember picture, overlay]
      \def\arraystretch{0.2}
      \node[left=15pt of e1](textofhere1){$e_1$};
      \draw[myarrow] (textofhere1) -- (e1);
      \node[right=25pt of e2](textofhere2){$e_2$};
      \draw[myarrow] (textofhere2) -- (e2);
    \end{tikzpicture}
\end{small}
\end{minipage}
\begin{minipage}{0.21\textwidth}
\begin{small}
\begin{gather*}
\begin{bmatrix} 
    x_{0,0}  & 1 & 1 & x_{0,3} \\
    x_{1,0} & 1 & 1 & \mypoint{e1}{\bm{{x_{1,3}}}} \\
    \makecmcolor{{x_{2,0}}} & \makecmcolor{x_{2,1}} & \mypoint{e2}{\makecmcolor{\bm{x_{2,2}}}} & \makecmcolor{x_{2,3}} \\
    1 & 1 & 1 & x_{3,3} \\
\end{bmatrix}
\end{gather*}
 \begin{tikzpicture}[remember picture, overlay]
      \node[right=15pt of e1](textofhere1){$e_1'$};
      \draw[myarrow] (textofhere1) -- (e1);
      \node[right=25pt of e2](textofhere2){$e_2'$};
      \draw[myarrow] (textofhere2) -- (e2);
    \end{tikzpicture}
\end{small}
\end{minipage}

    \item[Case 2] (b) has $(e, \insidegroup, \op)$.
    
    This case can be reasoned in a similar way as the previous case,
    by using \Cref{lemma:valid:insidegroup}.
    
    We construct a reduction instruction for (c) that expresses the same reduction.
    Suppose the reduction step in (b) covers parallelism factors $e_i,...,e_j$ on the reduction axis.
    Let $e'$ be the level corresponding to the parallelism factor right
    before $e_i$ row-wisely.
    Then $(e', \insidegroup, \op)$ is a desired reduction instruction.
        
    Below we give an example, with the reduction axis highlighted.
    
\noindent
\begin{minipage}{0.03\textwidth}
\quad
\end{minipage}
\begin{minipage}{0.18\textwidth}
\begin{small}
\begin{gather*}
\begin{bmatrix} 
    x_{0,0}  & 1 & 1 & 1 \\
    \mypoint{e1}{\bm{{x_{1,0}}}} & 1 & 1 & 1 \\
    \makecmcolor{x_{2,0}} & \makecmcolor{x_{2,1}} & \makecmcolor{x_{2,2}} & \makecmcolor{x_{2,3}} \\
    1 & 1 & 1 & 1 \\
\end{bmatrix}
\end{gather*}
 \begin{tikzpicture}[remember picture, overlay]
      \def\arraystretch{0.2}
      \node[left=15pt of e1](textofhere1){$e$};
      \draw[myarrow] (textofhere1) -- (e1);
    \end{tikzpicture}
\end{small}
\end{minipage}
\begin{minipage}{0.21\textwidth}
\begin{small}
\begin{gather*}
\begin{bmatrix} 
    x_{0,0}  & 1 & 1 & 1 \\
    x_{1,0} & 1 & 1 & \mypoint{e1}{\bm{{1}}} \\
    \makecmcolor{{x_{2,0}}} & \makecmcolor{x_{2,1}} & \makecmcolor{{x_{2,2}}} & \makecmcolor{x_{2,3}} \\
    1 & 1 & 1 & 1  \\
\end{bmatrix}
\end{gather*}
 \begin{tikzpicture}[remember picture, overlay]
      \node[right=15pt of e1](textofhere1){$e'$};
      \draw[myarrow] (textofhere1) -- (e1);
    \end{tikzpicture}
\end{small}
\end{minipage}

On the other hand, we can also show a counterexample of (b) $\geq$ (c),
where in the following hierarchy, 
$(e', \insidegroup, \op)$ is a valid reduction in (c) but there is
no way in (b) that can simulate the same reduction.

\noindent
\begin{small}
\begin{gather*}
\begin{bmatrix} 
    x_{0,0} & x_{0,1} & x_{0,2} & x_{0,3} \\
    x_{1,0} & x_{1,1} & x_{1,2} & \mypoint{e1}{\bm{{x_{1,3}}}} \\
    \makecmcolor{{x_{2,0}}} & \makecmcolor{x_{2,1}} & \makecmcolor{{x_{2,2}}} & \makecmcolor{x_{2,3}} \\
    1 & 1 & 1 & 1  \\
\end{bmatrix}
\end{gather*}
 \begin{tikzpicture}[remember picture, overlay]
      \node[right=15pt of e1](textofhere1){$e'$};
      \draw[myarrow] (textofhere1) -- (e1);
    \end{tikzpicture}
\end{small}

Similar counterexamples can also be shown for other cases.

\item[Case 3]
    (b) has $(e_2, \master{e_1}, \op)$.
    
    This case can be reasoned in a similar way as the previous case,
    by using \Cref{lemma:valid:master}.

    We construct a reduction instruction for (c) that expresses the same reduction. The choice of $e_1'$ and $e_2'$ is the same as Case 1.
    Then $(e_2', \parallel{e_1'}, \op)$ is a desired reduction instruction.
        
    Below we give an example, with the reduction axis highlighted.
    
\noindent
\begin{minipage}{0.03\textwidth}
\quad
\end{minipage}
\begin{minipage}{0.18\textwidth}
\begin{small}
\begin{gather*}
\begin{bmatrix} 
    x_{0,0}  & 1 & 1 & 1 \\
    \mypoint{e1}{\bm{{x_{1,0}}}} & 1 & 1 & 1 \\
    \makecmcolor{x_{2,0}} & \makecmcolor{x_{2,1}} & \makecmcolor{x_{2,2}} & \makecmcolor{x_{2,3}} \\
    1 & 1 & \mypoint{e2}{\bm{1}} & 1 \\
\end{bmatrix}
\end{gather*}
 \begin{tikzpicture}[remember picture, overlay]
      \def\arraystretch{0.2}
      \node[left=15pt of e1](textofhere1){$e_1$};
      \draw[myarrow] (textofhere1) -- (e1);
      \node[right=25pt of e2](textofhere2){$e_2$};
      \draw[myarrow] (textofhere2) -- (e2);
    \end{tikzpicture}
\end{small}
\end{minipage}
\begin{minipage}{0.21\textwidth}
\begin{small}
\begin{gather*}
\begin{bmatrix} 
    x_{0,0}  & 1 & 1 & 1 \\
    x_{1,0} & 1 & 1 & \mypoint{e1}{\bm{1}} \\
    \makecmcolor{{x_{2,0}}} & \makecmcolor{x_{2,1}} & \mypoint{e2}{\makecmcolor{\bm{x_{2,2}}}} & \makecmcolor{x_{2,3}} \\
    1 & 1 & 1 & 1 \\
\end{bmatrix}
\end{gather*}
 \begin{tikzpicture}[remember picture, overlay]
      \node[right=15pt of e1](textofhere1){$e_1'$};
      \draw[myarrow] (textofhere1) -- (e1);
      \node[right=25pt of e2](textofhere2){$e_2'$};
      \draw[myarrow] (textofhere2) -- (e2);
    \end{tikzpicture}
\end{small}
\end{minipage}

\end{description}
\qed

\subsection{Part3}
\label{sec:appendix:part3}

\begin{lemma}
\label{lemma:part3}
(d) $\geq$ (c).
\end{lemma}

\proof
Most reasoning is the same as \Cref{lemma:part2}.
For each case of a (c) reduction instruction,
we show how we construct the reduction instruction for (d) that
expresses the same reduction.
Remember that we attach a $(root, 1)$ to synthesis hierarchy (d).

\begin{description}
    \item[Case 1] (c) has  $(e, \insidegroup, \op)$.
      
    We construct a reduction instruction for (d) that expresses the same reduction.
    
    If the reduction covers the whole parallelism factor, then 
    (root, $\insidegroup, \op)$ is a desired reduction instruction.
    
\noindent
\begin{small}
\begin{gather*}
\begin{bmatrix} 
    x_{0,0}  & x_{0,1} & x_{0,2} & x_{0,3} \\
    x_{1,0} & \mypoint{e1}{\bm{{x_{1,1}}}} & 1 & 1 \\
    \makecmcolor{x_{2,0}} & \makecmcolor{x_{2,1}} & \makecmcolor{x_{2,2}} & \makecmcolor{x_{2,3}} \\
    1 & 1 & 1 & 1 \\
\end{bmatrix}
\end{gather*}
 \begin{tikzpicture}[remember picture, overlay]
      \def\arraystretch{0.2}
      \node[left=25pt of e1](textofhere1){$e$};
      \draw[myarrow] (textofhere1) -- (e1);
    \end{tikzpicture}
\end{small}

    If the reduction covers some (but not all) parallelism factors on the reduction axis, then $e$ itself is on the reduction aixs.
    
    Then $(e, \insidegroup, \op)$ is a desired reduction instruction.
    
\begin{small}
\begin{gather*}
\begin{bmatrix} 
    x_{0,0}  & x_{0,1} & x_{0,2} & x_{0,3} \\
    x_{1,0} & {x_{1,1}} & 1 & 1 \\
    \makecmcolor{x_{2,0}} & \mypoint{e1}{\bm{\makecmcolor{x_{2,1}}}} & \makecmcolor{x_{2,2}} & \makecmcolor{x_{2,3}} \\
    1 & 1 & 1 & 1 \\
\end{bmatrix}
\end{gather*}
 \begin{tikzpicture}[remember picture, overlay]
      \def\arraystretch{0.2}
      \node[left=25pt of e1](textofhere1){$e$};
      \draw[myarrow] (textofhere1) -- (e1);
    \end{tikzpicture}
\end{small}

    Again, if the reduction does not cover any parallelism factors on the reduction axis, then it forms reduction groups of size 1, and thus
    this does not form a reduction and won't be generated by \toolname.
    
    Below we also show a counterexample of (c) $\geq$ (d),
    where in the following hierarachy, $(e, \insidegroup, \op)$ is a valid reduction in (d) but there is no way in (c)
    that can simulate the same reduction since $x_{3}$ can
    be arbitrary numbers.
    
\begin{small}
\begin{gather*}
\begin{bmatrix} 
    x_{0,0}  & x_{0,1} & x_{0,2} & x_{0,3} \\
    x_{1,0} & {x_{1,1}} & 1 & 1 \\
    \makecmcolor{x_{2,0}} & \mypoint{e1}{\bm{\makecmcolor{x_{2,1}}}} & \makecmcolor{x_{2,2}} & \makecmcolor{x_{2,3}} \\
    x_{3,0} & x_{3,1} & x_{3,2} & x_{3,3} \\
\end{bmatrix}
\end{gather*}
 \begin{tikzpicture}[remember picture, overlay]
      \def\arraystretch{0.2}
      \node[left=25pt of e1](textofhere1){$e$};
      \draw[myarrow] (textofhere1) -- (e1);
    \end{tikzpicture}
\end{small}

    \item[Case 2] (c) has  $(e_2, \parallel{e_1}, \op)$.
    
    Similar as the previous case, we construct a reduction instruction for (d) that expresses the same reduction.
    
    Suppose the reduction step in (c) covers parallelism factors $e_i,...,e_j$ on the reduction axis.
    Let $e_1'$ be the level in the synthesis hierarchy (d) right
    before $e_i$ row-wisely.
    If the reduction covers the whole parallelism factor, then $e'$ would be the root.
    Let
    $e_2'$ be $e_j$. Then $(e_2', \parallel{e_1'}, \op)$ is a desired reduction instruction.

In the following example,  $e_1'$ = root.

\noindent
\begin{minipage}{0.03\textwidth}
\quad
\end{minipage}
\begin{minipage}{0.18\textwidth}
\begin{small}
\begin{gather*}
\begin{bmatrix} 
    x_{0,0}  & x{0,1} & x_{0,2} & x_{0,3} \\
    x_{1,0} & \mypoint{e1}{\bm{{x_{1,1}}}} & 1 & 1 \\
    \makecmcolor{x_{2,0}} & \makecmcolor{x_{2,1}} & \makecmcolor{x_{2,2}} & \makecmcolor{x_{2,3}} \\
    1 & 1 & \mypoint{e2}{\bm{1}} & x_{3,3} \\
\end{bmatrix}
\end{gather*}
 \begin{tikzpicture}[remember picture, overlay]
      \def\arraystretch{0.2}
      \node[left=20pt of e1](textofhere1){$e_1$};
      \draw[myarrow] (textofhere1) -- (e1);
      \node[right=25pt of e2](textofhere2){$e_2$};
      \draw[myarrow] (textofhere2) -- (e2);
    \end{tikzpicture}
\end{small}
\end{minipage}
\begin{minipage}{0.21\textwidth}
\begin{small}
\begin{gather*}
\begin{bmatrix} 
    x_{0,0}  & 1 & 1 & x_{0,3} \\
    x_{1,0} & x_{1,1} & 1 & 1 \\
    \makecmcolor{{x_{2,0}}} & \makecmcolor{x_{2,1}} & {\makecmcolor{{x_{2,2}}}} & \mypoint{e2}{\makecmcolor{\bm x_{2,3}}} \\
    1 & 1 & 1 & x_{3,3} \\
\end{bmatrix}
\end{gather*}
 \begin{tikzpicture}[remember picture, overlay]
      \node[right=25pt of e2](textofhere2){$e_2'$};
      \draw[myarrow] (textofhere2) -- (e2);
    \end{tikzpicture}
\end{small}
\end{minipage}
    
    \item[Case 3] (c) has  $(e_2, \master{e_1}, \op)$.
    
    This case is exactly the same as the case for $\parallelno$.

\end{description}
\qed

\end{document}